\documentclass[fleqn,usenatbib]{mnras}

\usepackage[T1]{fontenc}

\DeclareRobustCommand{\VAN}[3]{#2}
\let\VANthebibliography\thebibliography
\def\thebibliography{\DeclareRobustCommand{\VAN}[3]{##3}\VANthebibliography}

\usepackage{graphicx}	
\usepackage{amsmath}	
\usepackage{amssymb}	

\usepackage{ulem}	

\title[Rotation and Magnetic Activity on Kepler-411]
{
Starspot evolution, Differential Rotation and Correlation between Chromospheric and Photospheric Activities on Kepler-411
}

\author[F. Xu et al.]{
Fukun Xu%
$^{\href{https://orcid.org/0000-0002-8618-3551}{\includegraphics[width=8px]{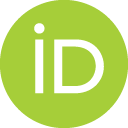}}}$%
$^{1,2}$\thanks{E-mail: \href{mailto:xufukun@ynao.ac.cn}{xufukun@ynao.ac.cn}, \href{mailto:shenghonggu@ynao.ac.cn}{shenghonggu@ynao.ac.cn}},
Shenghong Gu$^{1,2,3}$\footnote[1]{E-mail: shenghonggu@ynao.ac.cn},
Panogiotis Ioannidis$^{4}$ 
\\
$^{1}$Yunnan Observatories, Chinese Academy of Sciences, Kunming 650216, China\\ 
$^{2}$Key Laboratory for the Structure and Evolution of Celestial Objects, Chinese Academy of Sciences, Kunming 650216, China\\
$^{3}$School of Astronomy and Space Science, University of Chinese Academy of Sciences, Beijing 101408, China\\
$^{4}$Hamburger Sternwarte, Universit{\"a}t Hamburg, Gojenbergsweg 112, 21029 Hamburg, Germany\\
}

\date{Accepted 2020 December 4. Received 2020 December 4; in original form 2020 July 16}

\pubyear{2020}

\begin{document}
\label{firstpage}
\pagerange{\pageref{firstpage}--\pageref{lastpage}}
\maketitle

\begin{abstract}
We present an analysis of the starspot evolution, the surface differential rotation (SDR), the correlation between chromospheric activity indicators and the spatial connection between chromospheric and photospheric activities on the active star Kepler-411, using time series photometry over 4 years from Kepler, and spectroscopic data from Keck I 10-m
and Lijiang 2.4-m telescopes.
We constructed the light curve by re-performing photometry and reduction from the Target Pixel Files and Cotrending Basis Vectors with a manually redefined aperture using the software PyKE3.
An efficient program, GEMC\_LCM, was developed to apply a two-spots model to chosen light curve segments with three spot groups at fixed latitudes (30$\degr$, 45$\degr$), (30$\degr$, 60$\degr$) and (45$\degr$, 60$\degr$).
We found a periodic variation of the starspots at period of about 660 days
which independs 
on spot latitudes, and estimated the lower limit of SDR 
rate $\alpha = 0.1016(0.0023)$ and equatoral rotation period $P_{\text{eq}} = 9.7810(0.0169)$ days. 
Simultaneously, the relative variations of chromospheric activity indicators were derived by subtracting the overall mean spectrum from individual spectrum.
It is found that Ca II H and K emissions are strongly correlated with each other, and there also exists a correlation between H$\alpha$ and Ca II H \& K emissions, with large dispersion, in accordance with previous results. Furthermore we find the correlation between Ca II H and K emissions is different in 2011 and 2012.
%
The chromospheric emission variation shows a highly spatial anti-correlation with the light curve, suggesting a spatial connection between the chromospheric active region and spot region.
\end{abstract}

\begin{keywords}
stars: rotation -- starspots -- stars: chromospheres -- stars: magnetic fields
\end{keywords}

\section{Introduction}

Starspots are surface manifestations of tubes of magnetic flux, where the magnetic fields are strong enough to suppress the overturning
convection and thus block the energy from the stellar interior to the surface,
causing the spots to remain cooler (and therefore darker) than surrounding photosphere \citep{Bouvier1986, Strassmeier2009, Walkowicz2013}. 
Taking into account the rotation of stars, the starspots cause quasi-periodic modulations up to a few percent in the disc integrated light curve (LC) \citep{Valio2017}, that make themselves to be good tracers in measuring the stellar rotations.

Differential rotation (hereafter DR) on cool active stars is believed to play a critical role in the convective zones, as the case in the Sun.
It is expected to wind up magnetic field lines and to maintain a Solar dynamo \citep{Isik2011}, and plays an important ingredient in the theoretical models of the dynamo processes inside the convective zone, i.e. the so-called $\Omega$ effect that
"converts a poloidal magnetic field into a toroidal one by twisting it around the rotation axis" \citep{Petit2002}, 
based on the interaction between convective motion and (differential) rotation, although the detailed physical processes are still under investigation. 
The properties of DR can constrain the large-scale magnetic morphology of active stars and will provide some clues of the dynamo process \citep{Brandenburg2002, Olah2003},
thus can provide good observational constraints on theoretical dynamo models, and give us the opportunity to understand the relative phenomena such as activity cycle, etc.

The surface DR (SDR) can be easily measured if the rotation periods and latitudes of starspots are known.
Using Doppler Imaging (DI) \citep{Berdyugina1998b, Berdyugina2005} one can simultaneously trace series of spots on the photosphere of active stars, then detect SDR by comparing complex surface maps using cross-correlation method \citep{Donati1997, Kovari2012, Xiang2020}.
However DI is feasible only for bright and rapid rotators and requires continuous observations among adjoint rotations which largely restrict its utilization.
On the other hand, photometry of spotted stars is complementary to spectroscopic mapping technique, and may be able to derive SDR rates for an ensemble of active stars \citep{Reinhold2013}, because it is usually much more feasible for continuous observations on fainter targets than spectroscopy.

There are several methods for detecting SDR in photometry: seasonal period variations, Fourier method, light curve inversion and light curve modeling (LCM) \citep{Berdyugina2005}.
Method like the LCM \citep{Budding1977, Dorren1987}, which simulates the LC modulation by assuming a small number of circular spots, was proved to be an important tool in inversing the spot distribution, and applied to measure SDR on active rotators if observations covered several rotations are available \citep{Croll2006}.

Constraining the latitude of spot is usually not easy from one-dimensional LC, neither does SDR\citep{Lanza2016}.
One interesting technique based on analysis of spot-crossing events where spot is occulted by transiting planet was proven to be capable of solving this issue \citep[e.g.][]{2011ApJS..197...14D, 2012MNRAS.422L..72L, 2013ApJ...775...54S}.
This phenomenon was mentioned on Kepler-411 by \citet{Sun2019} and such an analysis is beyond the scope of this paper.

In addition, the magnetic field of the active regions on the surface of photosphere or in the low stellar atmosphere transports energy into the chromosphere, and produces emissions among series of chromospheric activity indicators \citep{Reinhold2017}, for example, the Ca II H \& K resonance and H$\alpha$ lines.
Many studies demonstrated tight correlations among observed emissions of different spectral diagnostics.
The variations of chromospheric activity indicators were also found to be co-located with the photometric variations under the rotational modulation \citep{Ash2020}.

In recent years, the high-precision space-borne photometric missions such as MOST \citep{Walker2003}, CoRoT \citep{Auvergne2009}, and most recent Kepler \citep{Koch2010} have made it possible to measure SDR. The extensive photometric observations represent an exciting opportunity to test and refine our understanding on stellar rotation.

The space-based Kepler mission \citep{Borucki2010} was primarily designed to detect Earth-sized and smaller planets around the habitable zones by using the transit-method.
Observing the same field more than 4 years, (May 2, 2009 - May 11, 2013), the Kepler telescope has provided invaluable data for about 150000 stars, which makes it a successful planet hunter and simultaneously a nearly perfect stellar variability observation machine, providing important science results across a diverse set of stellar phenomena.
One can thus try to exploit the long baseline and high precision Kepler observations to investigate the stellar activities.
For example, researchers have studied stellar rotation using starspot modulation for tens of thousands of stars \citep{Reinhold2013,Davenport2015}, and the SDR is also measured for thousands of stars \citep{Reinhold2013}.
One can also try to exploit 
Kepler LCs for inversing the spot distributions at different observing time, and therefore to estimate the potential temporal evolution of SDR.
This provides us the opportunity for the investigation of such kind of phenomenon on more active stars other than the Sun, AB Dor, etc.

In this paper we present a detailed analysis of surface inhomogeneities on the K2-type main sequence star Kepler-411 (KIC 11551692), using high-precision Kepler LC spanning over four years, to derive starspots and their evolution, and SDR at different epochs. 
In the section \ref{sec:obs}, we introduce the target, data reduction, and spectroscopic observations.
We describe our method for spot modeling and the method for measuring relative EWs from spectroscopic data in section \ref{sec:model}.
We then give the results and the possible explanation in section \ref{sec:rslt}, and finally the conclusions are summarized in section \ref{sec:concl}.
\section{Observations and Data Reduction}\label{sec:obs}
Kepler-411 has a magnitude of V = 12.55, a radius of 0.79 Solar radii and mass of 0.83 Solar mass\footnote{\label{www:411_info_kepler}\url{https://exoplanetarchive.ipac.caltech.edu}}. The relative parameters were greatly summarized on websites$^{\ref{www:411_info_kepler},}$\footnote{\label{www:411_info_exofap}\url{https://exofop.ipac.caltech.edu/kepler/edit_target.php?id=1781}}$^{,}$\footnote{\label{www:411_info_simbad}\url{http://simbad.u-strasbg.fr/simbad/sim-basic?Ident=Kepler+411&submit=SIMBAD+search}} 
and some of them are collected in table \ref{tab:411_info}.
The photometric rotation period as noted in previous studies of Kepler light curves is around 10.4 days \citep{McQuillan2013, Reinhold2013, Mazeh2015}
, and it was classified as a K2V main sequence star \citep{Frasca2016, Huber2014}.
\begin{table}
    \caption{Stellar parameters of Kepler-411}
    \label{tab:411_info}
    \begin{tabular}{ll}
        \hline
        Parameter           & Value \\
        \hline
        (RA, DEC)           & (19h10m25.347s, +49\degr31'23.712")$^{\ref{www:411_info_kepler}}$                  \\
        V mag               & 12.55$^{\ref{www:411_info_kepler}}$, 12.617$^{\ref{www:411_info_exofap}}$               \\
        Kepler mag          & 12.231$^{\ref{www:411_info_kepler}}$                                               \\
        Radius ($R_{\sun}$) & 0.79$^{\ref{www:411_info_kepler}}$, 0.75$^a$                           \\
        Mass ($M_{\sun}$)   & 0.83$^{\ref{www:411_info_kepler}}$, 0.81$^a$                          \\
        $T_{\text{eff}}$ (K)& 4974$^b$ , 4906$^a$, 4920$^{\ref{www:411_info_exofap}}$     \\
        $[$Fe/H$]$   & 0.07$^{\ref{www:411_info_exofap}}$                                                 \\
        log(g)              & 4.54$^b$, 4.60$^a$, 4.620$^{\ref{www:411_info_exofap}}$    \\
        Inclination (\degr) & 88.8$^c$                                                         \\
        v \text{sin}(i) (km/s)     & 3.64$^a$                                                        \\
        \hline
        \multicolumn{2}{l}{$^a$ \citep{Morton2016}  }\\
        \multicolumn{2}{l}{$^b$ \citep{Wang2014}    }\\
        \multicolumn{2}{l}{$^c$ \citep{Morton2014}  }\\
    \end{tabular}
\end{table}

\subsection{Photometric data preparation}

The space-borne Kepler telescope \citep{Borucki2010} provides amazing amount of photometric data which is delivered in 18 quarters (Q0 - Q17), each spanning about 90 days, with exception for the commissioning phase Q0 ($\sim$ 10 days), Q1 ($\sim$ 33 days) and Q17 ($\sim$ 33 days, observations were interrupted due to failure of the reaction wheel 4). Another terrible event during the observations is the permanent failure of Module 3 since 17:52 UTC on Jan 9, 2010 (Q4), which led to 20\% of the FOV to suffer one-quarter data outage every year as Kepler performed its quarterly rolls \citep{KDCH}.
The data of Kepler-411 span 15 quarters (0 - 3, part of 4, 5 - 7, 9 - 11, 13 - 15, 17) for long cadence and 5 quarters (11, 13 - 15, 17) for short cadence. The absence in Q8, Q12, Q16 and part of Q4 results from the Module 3 failure.

Kepler data have been processed by three versions of pipeline so far, which was designed to detect planetary signals, starting with the elementary Presearch Data Conditioning (PDC) pipeline. This pipeline was changed to the so-called PDC-MAP (Maximum A Posteriori) pipeline \citep{Twicken2010a, Stumpe2012, Smith2012} because the PDC pipeline only coarsely removed stellar variability signals. Recently, all Kepler data have been reprocessed by the PDC-msMAP (multiscale MAP) pipeline, which applies an "overcomplete discrete wavelet transform" to correct LC in several frequency bands and thereby allows a better performance in removal of the systematics \citep{Stumpe2014}.
However, the stellar variabilities usually exceed a non-unique and much broader range of periodic durations and variations than planetary transits, and could be ignored or misinterpreted by the pipelines. Thus it is usually recommended to do the systematic error corrections by hand according to the individual analysis in the investigation of stellar variabilities \citep{pyke3_1,pyke3_2}.

On the other hand, the autonomous determined pixel mask used in PDC works well for the searching for the planetary transit signals, but it is not always the case for the study of stellar activities because now we are mostly interested in the LC variations in multiple time scale, that can be neglected in the autonomous process. This neglect in the method is most apparent for variable stars due to the internal brightness variations, which are more important than transits in our case. The other disturbances come from the systematic errors such as the attitude tweaks, focus changes and the momentum desaturation \citep{KDCH}, which lead to variations of photon distributions on CCD.
It was found that, the pixel masks of almost all quarters in the autonomous determination of Kepler-411 are indeed improper for the investigation of stellar activity such as spots, more or less. 

As a result, it becomes necessary in our case to construct LCs from the initial calibrated target pixel file (TPF, which contains the original pixel level photometric data, available at Mikulski Archive for Space Telescopes (MAST) at STSci data base\footnote{\label{www:MAST}\url{https://archive.stsci.edu/kepler/}}) with re-defined apertures, and re-performing the data reduction to remove systematic errors using cotrending basis vectors (CBV, also available at MAST$^{\ref{www:MAST}}$) and finally excluding remaining disturbances.
We used PyKE software of Version 3 \citep{pyke3_1,pyke3_2} as follows:
firstly task "keppixseries" was used to check if there is any flux contamination in the target LC by considering the TPF for each quarter, and task "kepmask" was used to mark the optimal aperture for re-performing photometry; then the LCs were extracted by using task "kepextract" for questionable quarters; finally the systematic errors were removed through the fitting of CBVs by using task "kepcotrend".

The TPFs consist of "postage stamp" snapshots of the object taken at every time record, which can be used to check if there is any flux contamination in the extracted LC attributed to improperly determined apertures.
During inspections, we encountered PDC-pipeline specific misinterpretations in almost all quarters, especially in Q2, Q10 and Q14, because the apertures chosen were too small to reveal the stellar intrinsic variations and a substantial proportion of pixels was abandoned.
This might not bring bad consequences for hunting exoplanet transits, however will lead to not only the notable loss of the luminance in different quarters, which appeared as apparent discontinuities between quarters and even could not be correctly rectified by PDC-pipeline, but also distortions on LC at long-time scale, which are usually difficult to be distinguished from the stellar intrinsic variations.
As a result, a careful choice of the optimal aperture with wider scale was necessary to include the intrinsic variations of the target as much as possible under the promise of high signal-to-noise ratio (SNR).

\begin{figure}
	\includegraphics[width=\columnwidth]{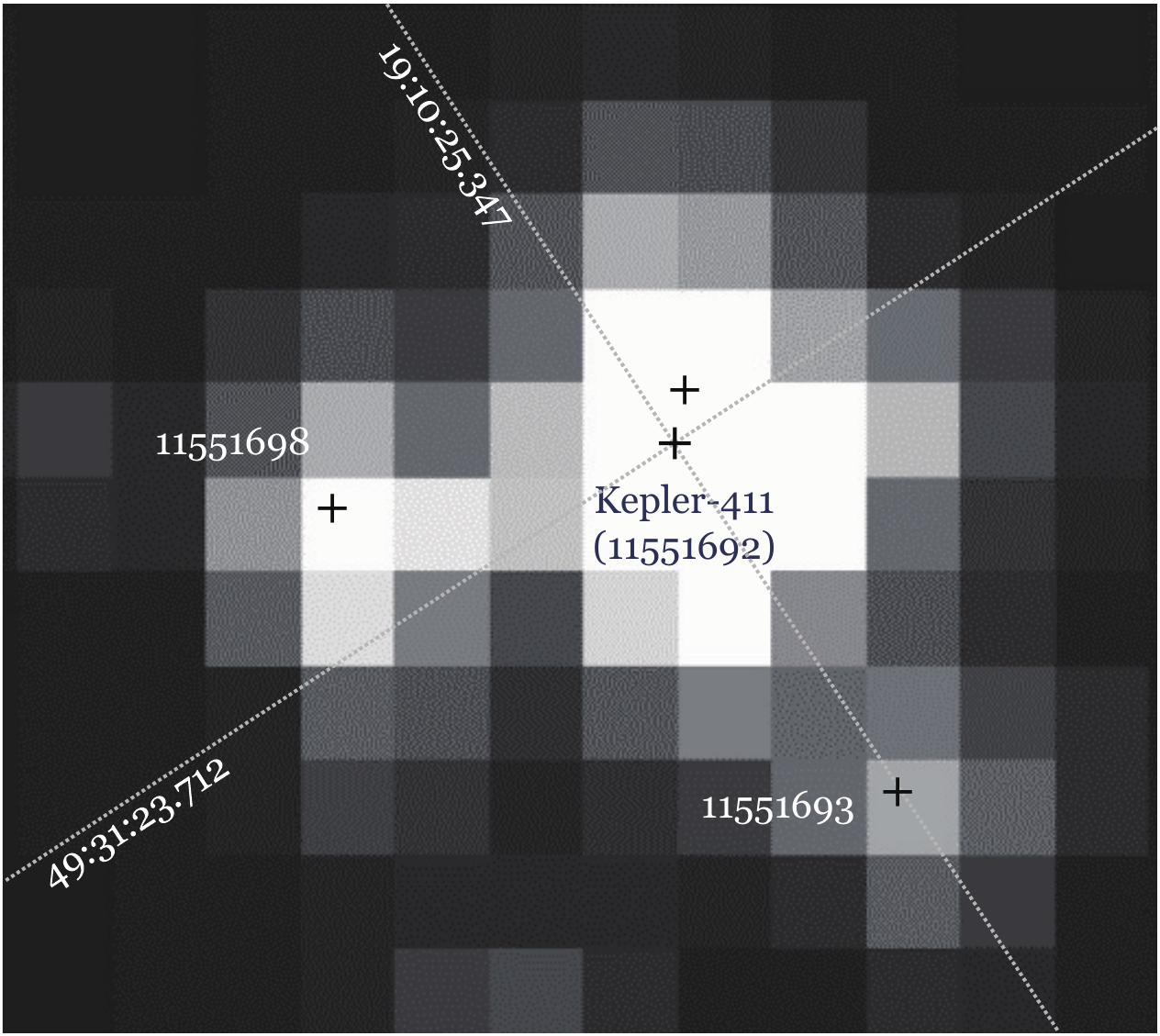}
	\caption{FFI of Kepler-411 in quarter Q2, neighbors are marked with solid crosses.}
	\label{fig:crowd_FFI}
\end{figure}

However, the aperture should not be too large to include the contributions of crowded neighbors within the FOV of TPF which might lead to questionable LCs especially for wide FOV telescopes,
figure \ref{fig:crowd_FFI} shows the crowded neighbors and the CCD figure from the full frame image (FFI) of Q2.
Two neighbors, KIC 11551693 and KIC 11551698, are much fainter and far enough from the target so that they can be easily excluded from the photometric aperture.
The third neighbor, which might be a nearby companion, with distance $3.449$ arcsec and $3.653$ mag fainter than Kepler-411$^{\ref{www:411_info_exofap}}$ \citep{Furlan2017}, is impossible to be excluded due to its proximity (even smaller than the image scale of Kepler, i.e. 3.89 arcsec per pixel).

The spacecraft performs a roll maneuver every three months to transmit data to the Earth, results in redistributions of the target flux across adjacent pixels on different CCDs. To compensate, the target photometry aperture is redefined after each roll maneuver.
Three apertures were chosen in photometry relative to the observing seasonal variances of the telescope after several tests. That is, Q0, Q1, Q5, Q9 and Q13 present almost the same distribution of the photons on the CCD panel for the Spring season of Kepler, Q2, Q6, Q10 and Q14 for the Summer season, and the left Q3, Q7, Q11 and Q15 for the Fall season, and note that the Winter quarters Q4 (partly), Q8, Q12 and Q16 are absent due to the failure of Module 3.

LCs were extracted using task "kepextract" from TPF with chosen apertures.
Such re-extracted LCs must be detrended for spacecraft induced features by removing the CBVs with custom weights determined using task "kepcotrend".
An order of 3 CBVs was chosen in "kepcotrend" after several tests, starting with two basis vectors and increasing the number of vectors monotonically until deciding upon a converged fitting among all quarters\footnote{\url{https://keplerscience.arc.nasa.gov/PyKEprimerCBVs.shtml}}.

Three kinds of data points were removed from the resultant LCs to prevent them from disturbing the following inversing procedure.
Data points marked as bad quality or questionable by Kepler \citep{KDRN, KDCH} were firstly removed. Such points are produced, for example, by the impact of high-energy particles, introducing sharp distortions to the LC.
The second was the points relative to the monthly "Earth-points" and the quarterly rolls, which were intended to be fully removed by the reduction, however there were some exceptions remained practically.
At last the non-spot caused effects such as the exoplanet transit events, flares, etc. were also excluded because we were presently interested in the stellar variation by spot modulation.

\begin{figure*}
	\includegraphics[width=0.95\textwidth]{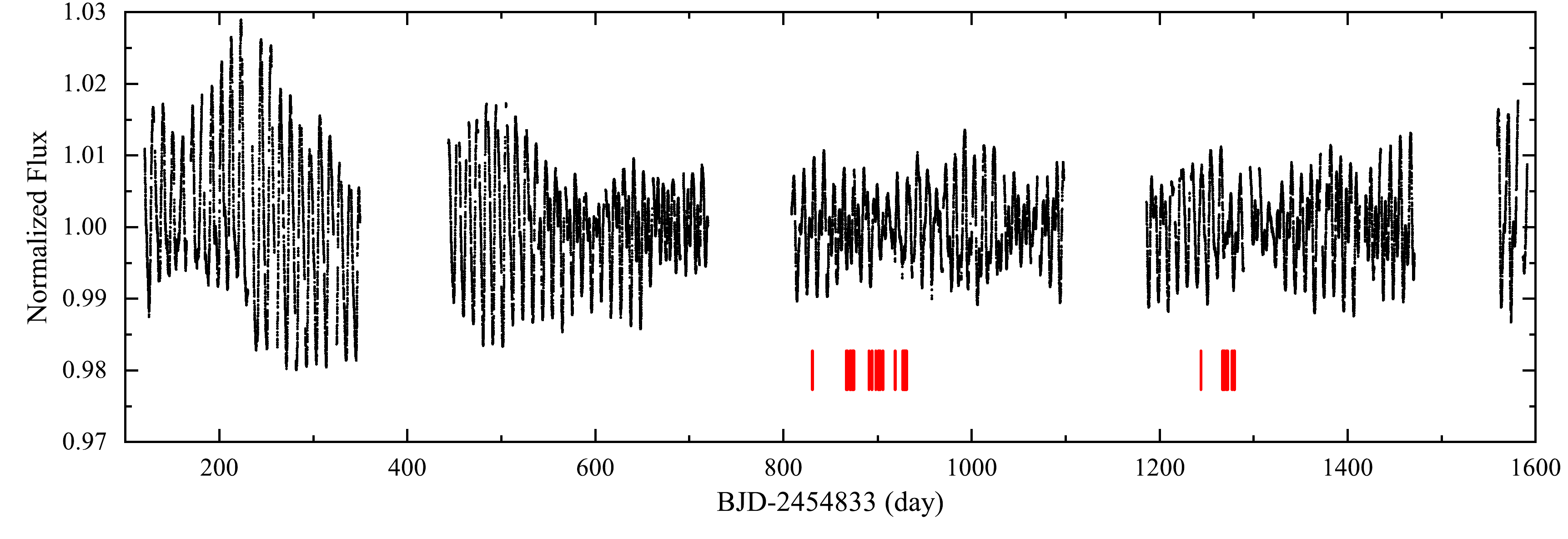}
	\caption{Resulting LC with our scheme. 
	The time when spectra were obtained from Keck/HIRES are marked with red vertical lines.}
	\label{fig:lc_all}
\end{figure*}

\begin{figure}
    \includegraphics[width=\columnwidth]{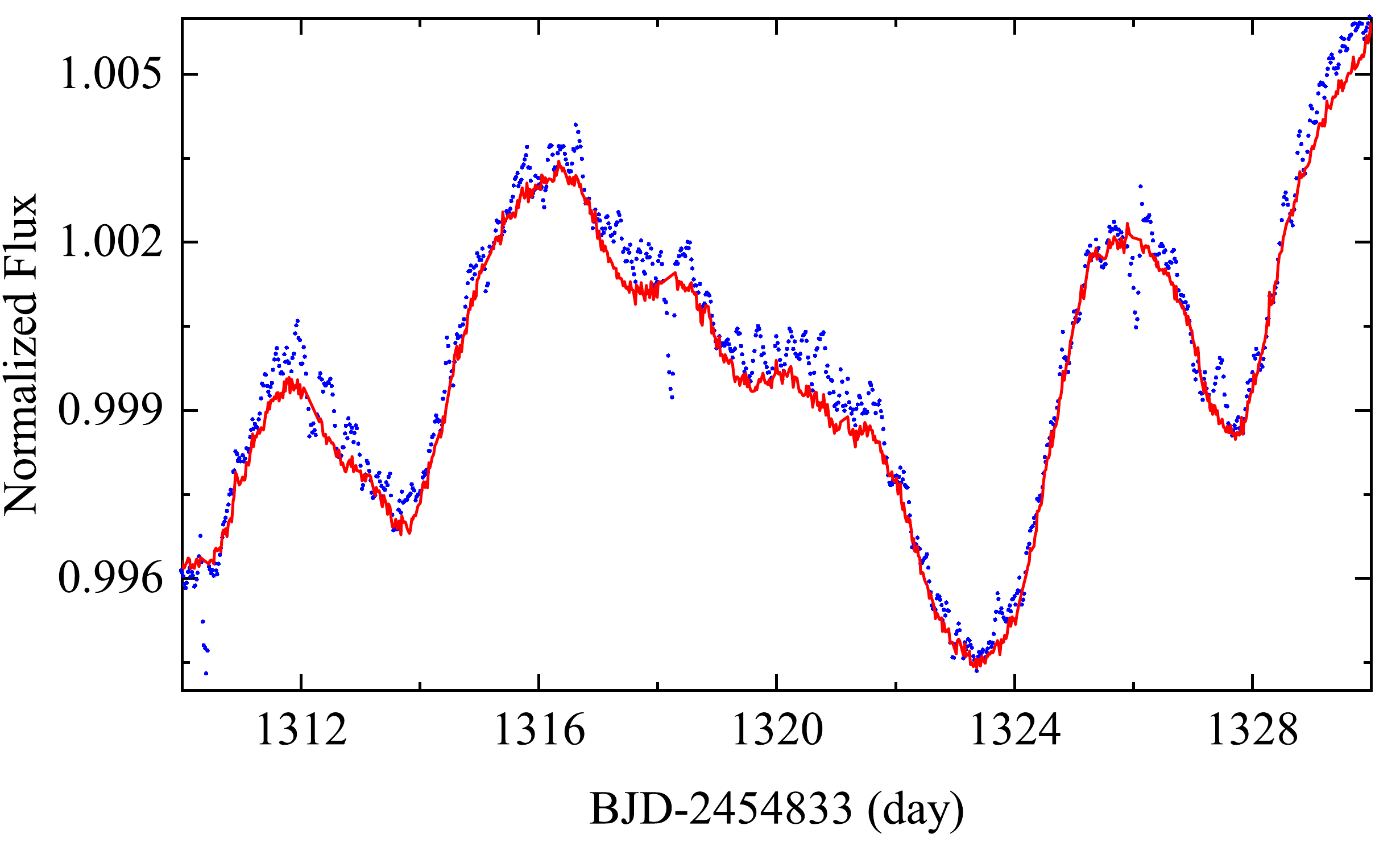}
    \caption{Sample LCs of Kepler 411 from PDC (blue dots, revealing high frequency noise), and our scheme (red curve), note that the exoplanet transit events have been removed in our case.}
    \label{fig:lc_q14_compare}
\end{figure}

Finally each quarterly LC was normalized and centered around unity by dividing the flux by its median value to minimize discontinuities of the measured fluxes between quarters \citep{Reinhold2017}.
Figure \ref{fig:lc_all} shows the resulting LC, for which the flux levels and the amplitudes have been corrected, and the useless points are removed. Here we use the PDC reduced data for Q0 and Q17 because their time length are too short to appropriately remove systematic errors using the same scheme as other quarters, and Q4 is ignored because it suffered from the Module 3 failure.
It can be found that the resulting LC are better for spot modeling than PDC LC, more or less, mainly depending on the difference between the autonomous determined and the revised optimal apertures.
The most notable improvement comes from Q14 as displayed in figure \ref{fig:lc_q14_compare}, where the PDC reduced LC reveals a prominent high frequency noise, this indicates a better extraction and reduction of the LC.

\subsection{Spectroscopy}
The spectroscopic data we used consist of two parts. 
One is the observations collected from the Keck Observatory Archive (KOA\footnote{\label{www:koa}\url{http://nexsci.caltech.edu/archives/koa/}}), which were carried out during 2011 - 2012 by the high-resolution echelle spectrograph of the Keck I 10-m telescope at Keck Observatory, Mauna Kea, Hawaii  \citep[Keck/HIRES,][]{Vogt1994}.
Keck/HIRES is an echelle spectrograph covering 3000 to 10000 Angstroms, with a spectral resolution of R = 67000.
The other is our additional observations on Dec, 2017 by the Yunnan Faint Object Spectrograph and Camera attached to the Lijiang 2.4-m Telescope at the Lijiang station of Yunnan Observatories, CAS \citep[Lijiang/YFOSC,][]{Fan2015, Wang2019}.
A $2408 \times 4096$ pixel CCD camera, with a pixel size of $13.5 \mu m^2$, was used along with the spectrograph. Using the echelle mode of YFOSC, we were able to record spectra between 3400 and 10000 Angstroms in 13 echelle orders, which provides a spectral resolution of $R= 2500$ around 6500 Angstroms.

The reduction of the spectra was carried out using standard tasks in the IRAF package\footnote{IRAF is distributed by the National Optical Astronomy Observatories, which is operated by the Association of Universities for Research in Astronomy (AURA), Inc, under cooperative agreement with the National Science Foundation.}, 
which basically includes bias 
correction, flat-field division, scattered light correction, and extraction of spectra from echelle orders. 
The wavelength calibration was done using the spectrum of arc lamp, which was taken in the beginning of the night or just after each exposure of the target.
Totally 31 spectra of Keck/HIRES (i.e. 21 spectra in 2011, 10 spectra in 2012) and 7 spectra of Lijaing/YFOSC are obtained (table \ref{tab:spec_obs_info}).

\begin{table}
    \caption{Spectroscopic observations at Keck 10-m telescope and Lijiang 2.4-m telescope}
    \label{tab:spec_obs_info}
    \begin{tabular}{lrl}
        \hline
        Year & Number of observations & Instrument \\
        \hline
        2011 & 21 & Keck/HIRES \\
        2012 & 10 & Keck/HIRES \\
        2017 & 7  & Lijiang/YFOSC \\
        \hline
    \end{tabular}
\end{table}

\section{Model and Analysis}\label{sec:model}

\subsection{Spot model} \label{subsec:gemc_lcm}

\subsubsection{LCM}
As a model towards photometric observations, 
LCM \citep{Budding1977, Dorren1987} was proposed to analytically simulate the combined photometric LC by assuming a small number of circular spots covering on the stellar photosphere, which might help to reduce the parameter space to some extent.
\citet{Budding1977} estimated the drop in light intensity caused by one circular spot by defining the "$\sigma$-integrals" for covered area projected on the line of sight as a fundamental formula,
\begin{equation}
    \label{eq:lcm_sigm_mn}
    \sigma_m^n = 1/\pi \int\int_{\text{spot\ area}}{x^m z^n dxdy}
\end{equation}
where $\sigma_0^0$ represents the projected area of the spot coverage, while $\sigma_1^0$ represents the effect of linear limb darkening, and the $xyz$ Cartesian coordinate system orients the z-axis towards the observer, which can be related to the spherical polar system of spot longitude $\lambda$ and latitude $\beta$, and hence the "$\sigma$-integrals" becomes $\sigma_m^n \equiv \sigma_m^n(i, \lambda, \beta, \gamma)$, where $i, \gamma$ and following used parameters in this model are listed in Table \ref{tab:lcm_prmt}.
Then the theoretical light intensity as a function of time, $I_c(t)$, due to the effect of all nonoverlapping uniform circular spots, is
\begin{equation}
    \label{eq:lcm_ic}
    I_c(t) = U \left[1-\sum_j^{N_{\text{spots}}}{(1-\kappa_{w_j}) {3\over{3-u}} [(1-u)\sigma_0^0+u\sigma_1^0]}\right]
\end{equation}
Here the $j$ subscript represents the $j^{\text{th}}$ spot. 

Practically $i, U, u$ and $\kappa_w$ can be taken as constants and common to all spots.
The geometry of one circular spot can be described by three parameters, i.e. latitude, longitude and radius. 
Longitude $\lambda$ is an ambiguous concept when stellar rotation is considered and can be defined better as a function of time $t$ by the rotation period $P$, epoch $E$ and longitude $\lambda(0)$ when spot appears at $\lambda(t) = 2\pi(t-E)/P-\lambda(0)$.
Thus the number of independent parameters for each spot is 4, because $E$ and $\lambda$ are usually relevant so that either of them can be set to $0$ for convenience.
\begin{table}
    \centering
    \caption{Input parameters in LCM}
    \label{tab:lcm_prmt}
    \begin{tabular}{ll}
        \hline
        Parameter        &   Definition                                              \\
        \hline
        $i$ ($\degr$)       &   Inclination of stellar rotation axis to line of sight   \\
        $U$               &   Stellar unspotted intensity                             \\
        $u$               &   Linear limb-darkening coefficients ($0 \sim 1$)              \\
        $\kappa_w$        &   Spot-to-photosphere intensity ratio ($0 \sim 1$)             \\
        $P$ (days)        &   Spot period                                             \\
        $E$               &   Spot epoch                                              \\
        $\lambda$ ($\degr$) &   Spot longitude ($0 \sim 360$)                                \\
        $\beta$ ($\degr$)   &   Spot latitude ($-90 \sim 90$)                                \\
        $\gamma$ ($\degr$)  &   Spot angular radius ($0 \sim 90$)                            \\
        \hline
    \end{tabular}
\end{table}


\subsubsection{GEMC\_LCM}

Spot modeling to reconstruct a two-dimensional map of the stellar surface from one-dimensional photometric time series is a well-known illposed problem \citep{Lanza2016}.
One reason is that one usually has to deal with a large parameter space in inversion or modeling, which blocks the algorithm converging to the true solution during iteration. The other is due to the weak constraints from the observation itself, which makes the optimization algorithm much easier to fall into local minimal solutions due to even small disturbances such as the noise.

When it comes to the optimization algorithm, \citet{Tregloan-Reed2013, Tregloan-Reed2015} developed an efficient program, so-called genetic evolution Markov chain (GEMC), which consists of two steps: a hybrid between Monte Carlo Markov Chain \citep[MCMC,][]{Press2007} and genetic algorithm \citep[GA,][]{Charbonneau1995}, and a parameter estimation based on the differential evolution Markov Chain (DE-MC) proposed by \citet{Braak2006}.
By combining both superior global optimization power of GA and high efficiency on parameter space exploration of MCMC, GEMC is capable to jump large distances across the solution space and thus more likely to find the global solution for simulating high-dimension, multi-modal systems \citep{Gregory2011, Sun2017}. 

A program, named "GEMC\_LCM", was developed to find the plausible configuration of starspots from an analytical fitting of model to a given LC with efficiency.
It was designed to employ a small parameter space by using LCM \citep{Budding1977} and to simultaneously inherit the superior global optimization power of GEMC.
The GEMC\_LCM was constructed using Fortran and Python\footnote{We employed and modified module "MCMC" by Marko Laine (\url{https://github.com/mjlaine/mcmcf90}) and "sort" by Samuel Ponce (\url{http://www.gnu.org/copyleft.gpl.txt}), and compiled by f2py and gfortran with OpenMP.}, and devoted to improve the computational efficiency.
The optimization was done using the $\chi^2$ minimization method.

\subsection{LC and SDR analysis}\label{subsec:sdr}
The approach to SDR measurement on Kepler-411 includes several steps.
Firstly several LC segments with two dominate spots were chosen, a two-spot model with fixed latitudes was employed, assuming no evolution for the spot distributions besides linear evolution of their radii.
Then the shears of the SDR were calculated for all segments by assuming a constant rotation period of the equator, therefore the temporal variations of the SDR were derived.

\subsubsection{LC analysis}

In order to derive a meaningful model we need to select a time span, which is both long enough to provide an appropriate visibility of the stellar surface, and short enough to minimize the effects of spot evolution.
During the analysis, the LC is split into equally sized intervals covering the time span of 21.5 days, which is slightly longer than twice of the rotation period (i.e. about 10.4 days).

It is always not trivial to fit the LC variations with spot modeling, due to both the large parameter space in optimization and the weak constraints on spot latitude in photometry, each of which usually leads to highly degenerate solutions in practice. Indeed a blind test for the reliability of the SDR detection suggested that DR studies based on full-disc LC alone need to be treated with caution \citep{Aigrain2015}.
A small number of spots is helpful to reduce the parameter space, and to increase the probability of the optimization algorithm to converge to the global minimum.

Synthesized LCs using the same fixed parameters for Kepler-411 were created to check the global optimizing capability of models with two and three spots, respectively.
The modeling was done with both the exact and deviated values of fixed parameters, for example the spot-to-photospheric intensity ratio, as input. 
Both models converged to the ``true'' solution in almost all runs when the exact fixed parameters were used, only the two spot model revealed a high probability to converge to the local minimum near the ``true'' solution if the fixed parameters were slightly deviated.
On the other hand, fitting the real LC with three spot model resulted in families of solutions with equally good fits which reduced its reliability, 
while the one with two spot model could usually converge to unique and stable solution, except the case with a ``flip'' between spot latitudes \citep{Davenport2015}.
So we only try to fit the LC segments by assuming two long-lived circular spots.
By running the code many times, we could manage to choose the group of solutions with smaller residuals as the final result by comparing consistence among all segments. 

The LC segments were picked by eyes among the whole Kepler dataset, which fulfill above conditions and show no spot evolution, listed in table \ref{tab:lc_chosen}. About two-thirds of the era were sparsely covered from Q5 to Q15, it is difficult to include the beginning quarters Q0 - Q3 because more than 3 spots are always visible during this time.
\begin{table}
    \caption{LC segments of Kepler-411 picked for modeling.}
    \label{tab:lc_chosen}
    \begin{tabular}{rrrrl}
        \hline
        Segment & Start-epoch & End-epoch & Mid-epoch & Quarter \\
        No. &  \multicolumn{3}{|c|}{BJD-2454833}  & No. \\
        \hline
        1   &   480.0   &   501.5   &   490.75  &   Q05     \\
        2   &   542.0   &   563.5   &   552.75  &   Q06     \\
        3   &   632.0   &   653.5   &   642.75  &   Q07     \\
        4   &   695.0   &   716.5   &   705.75  &   Q07     \\
        5   &   852.5   &   874.0   &   863.25  &   Q09     \\
        6   &   1004.0  &   1025.5  &   1014.75 &   Q11     \\
        7   &   1075.0  &   1096.5  &   1085.75 &   Q11     \\
        8   &   1190.0  &   1211.5  &   1200.75 &   Q13     \\
        9   &   1332.0  &   1353.5  &   1342.75 &   Q14     \\
        10  &   1383.0  &   1404.5  &   1393.75 &   Q15     \\
        11  &   1436.0  &   1457.5  &   1446.75 &   Q15     \\
        \hline
    \end{tabular}
\end{table}

The modeling simulates the star as a sphere with uniform surface brightness and limb darkening, on which circular spots with uniform temperature are put. The linear limb darkening coefficient $u$ was fixed to an approximate value of $0.7$ from the table of \citet{Claret2012, Claret2013} for a star with $T_\text{eff} = 4900\sim5000 K$ and $\text{log(g)} = 4.5$. The spot-to-photosphere intensity ratio $\kappa_w$ was fixed to $0.3$ ($\kappa_w \sim (T_{\text{spot}}/T_{\text{sphere}})^4$) with spot temperature $T_{\text{spot}} \sim 3700 K$ estimated from \citet{Berdyugina2005}. The stellar inclination $i=88.8\degr$ based on table \ref{tab:411_info}. The stellar surface brightness $U$ was taken as free parameter because it is usually unknown if the target is always occupied by spots as in our case.

A fundamental limitation for the inversion of spot latitude is no unique solution.
This becomes worse in the case of high inclination of the rotation axis due to the disturbance coming from the noise. For example, simulation by \citet{Ioannidis2016} revealed that for inclinations in excess of about 70$\degr$, the noise in the data exeeds the maximally possible deviations introduced by the equatorial spots for Kepler-210.
Another problem is that one can not even distinguish one semisphere from the other for the extremely high inclinations.

Considering that we also likely suffered from the same situation due to the high inclination of 88.8$\degr$, an alternate way was applied to assume the latitude distribution based on the so-called "thin-flux-tube-evolution model" simulation \citep{Granzer2002}, from which the allowed range of 30 - 60$\degr$ of the spot latitude distribution was obtained referring the rotational period of about 10.4 days and mass of about 0.8 $R_{\sun}$ (table \ref{tab:411_info}).
Furthermore, three cases of latitude combinations, (30$\degr$, 45$\degr$), (45$\degr$, 60$\degr$) and (30$\degr$, 60$\degr$), corresponding to the lower, medium and upper values of the spot latitude range, were chosen in the two-spot model to estimate the relative variations of the magnetic activity.
%
A linear evolution of spots' radii is also allowed to reveal the rapid variations among adjoint rotations.

\subsubsection{SDR}
On the Sun the SDR is observed in relative motion of sunspots and can be expressed by a quadratic law
\begin{equation}
    P(\beta)=P_{\text{eq}}/(1-\alpha \sin^2{\beta})
    \label{eq:DR}
\end{equation}
where $P(\beta)$ is the rotation period at latitude $\beta$, $P_{\text{eq}}$ is the equatorial rotation period.
Strength of the SDR can be quantified by the relative differential rotational rate $\alpha$ which is expressed as the ratio of the rotational shear $\Delta\Omega$ to the rotation rate $\Omega = 2\pi/P$ at the equator, $\alpha = \Delta\Omega/\Omega_{\text{eq}}$.
It can also be characterised by the lap time, $\tau_l = 2\pi/\Delta\Omega = P_{\text{eq}}/\alpha$, which is the time the equatorial region needs to lap the pole.
For instance, on the Sun with $\alpha = 0.2$, $\Delta\Omega = 0.055$ rad d$^{-1}$, and the lap time is 115 d.

A general way to investigate SDR on stars other than the Sun is to assume a similar formula of SDR law on them by analogy \citep{Henry1995,Kovari2012,Reinhold2015}, while a positive $\alpha$ represents a Solar-like DR, a negative $\alpha$ represents a anti-Solar DR and $\alpha = 0$ means the star rotates as a solid body.
Then the SDR can be measured from the rotation period and latitude of spot.

It is worthy to note that, we calculate SDR after accomplishment of spot modeling other than inverse it directly in modeling.
Because the variation of $P_{\text{eq}}$ or $\alpha$ will cause the relative modifications of rotation periods as well as latitudes of all spots, this makes optimization falling into lower efficiency and more difficult to converge to the plausible physical solutions in practice.

\subsection{Relative variations of chromospheric activity}

Chromospheric activity produces filled-in or emission in some strong photospheric lines, which are formed at different atmospheric heights, such as the 
Ca II H \& K and H$\alpha$ (formed in middle chromosphere)
lines, and are taken as proxies of chromospheric activity \citep{Eberhard1913}.
They have been proven to be very important and useful chromospheric activity indicators in the optical spectral range.

The most commonly used indicator of stellar magnetic activity is the ionized calcium H \& K line core emissions \citep{Wilson1968},
developed by the Mount Wilson Observatory (MWO) HK Project, or the derivative index $R_{\text{HK}}$, is approximately proportional to the square root of the mean magnetic field strength at the stellar surface \citep{Schrijver1989}.
H$\alpha$ line is also usually used as a chromospheric activity indicator, which is believed to be correlated with the Ca II H \& K indicators.

We selected the Ca II H \& K (3968 and 3933 A) and H$\alpha$ (6563 A) lines as tracers of chromospheric activity \citep{Montes2004, Zhang2015} which were extracted and measured from the observed spectra.
Comparing with the average level of chromospheric activity, we can estimate the relative chromospheric activity to some degrees besides checking the features of magnetic activity in the chromosphere. 


Noting that what we are interested in are the relative variations of the chromospheric activity at different times, we chose an alternative way to measure the relative equivalent width (EW) of the emission lines by subtracting an overall average spectrum from the spectra taken over all nights, to overcome the difficulty in finding the continuum \citep{Shkolnik2005} and to reduce the deviation 
introduced by choices of continuum among different spectra.
The approach was as follows.

Firstly, all spectra were transformed to the spectrum with the highest SNR, by using a robust nonlinear least squares curve fitting over the surrounding spectral portions around the emission lines with three parameters: a wavelength shift $\Delta\lambda$ by removing radial velocity deviation, a scaling factor $c$ and a linear trend $d$ at reference wavelength $\lambda_0$, i.e. $F_{\lambda}' = cF_{\lambda+\Delta\lambda}+d(\lambda-\lambda_0)$.
Here the Huber loss function \citep{Huber1964} for robust regression was used to suppress the contributions from abnormal points.

For Ca II H \& K lines, the very strong photospheric lines suppress the spectral continuum, making the consistent normalization difficult.
We chose narrow bandpasses centered on these two lines, which set a normalization level as described by \citet{Shkolnik2005} and \citet{Sun2017}. Two 5.4 Angstroms spectral ranges, centered on H and K lines respectively, were chosen to isolate the reversals while minimizing any apparent continuum differences induced by varying illumination of the CCD, and the surrounding spectra were obtained by assuming a typical width of 0.9 Angstroms for the emission cores.

For the H$\alpha$ line, however, the continuum is easy to find and the surrounding spectral windows were chosen as 6557.5 - 6560 and 6564.44 - 6566.94 Angstroms to exclude the Earth atmospheric water absorption lines (6557.2, 6560.5 and 6564.2 Angstroms).

Secondly, 
an average spectrum was derived from all observed spectra, 
then transformed to unity.
And then all spectra were normalized to unity using the same transformation as above, to suppress the possible disturbances coming from the errors in finding the continuum for individual spectrum.
The normalization of the average spectrum to unity was done by fitting a straight line to the edges of the spectra portions for the Ca II H \& K lines, whereas by directly dividing by the continuum for the H$\alpha$ line.

\begin{figure}
    \includegraphics[width=\columnwidth]{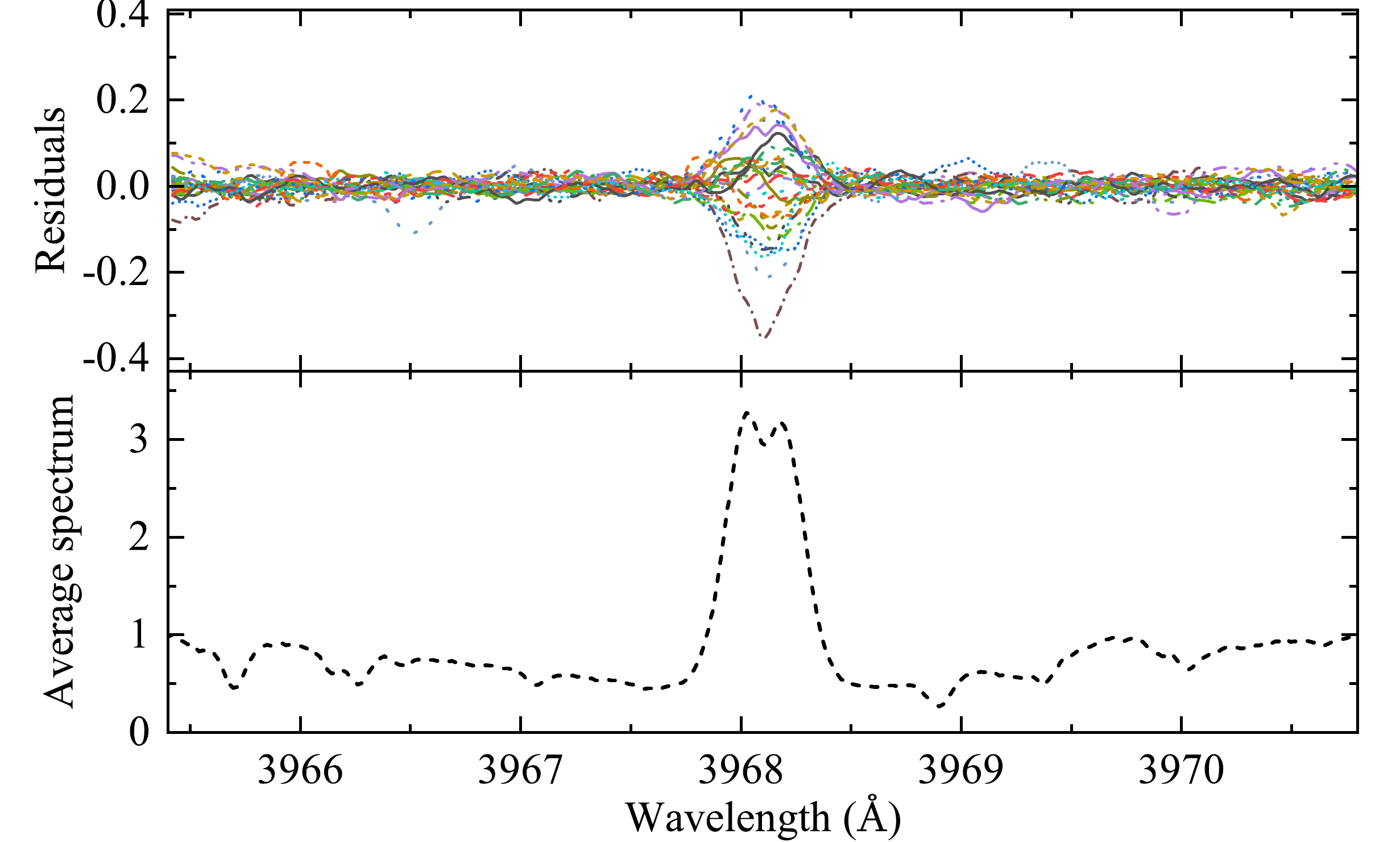}
    \includegraphics[width=\columnwidth]{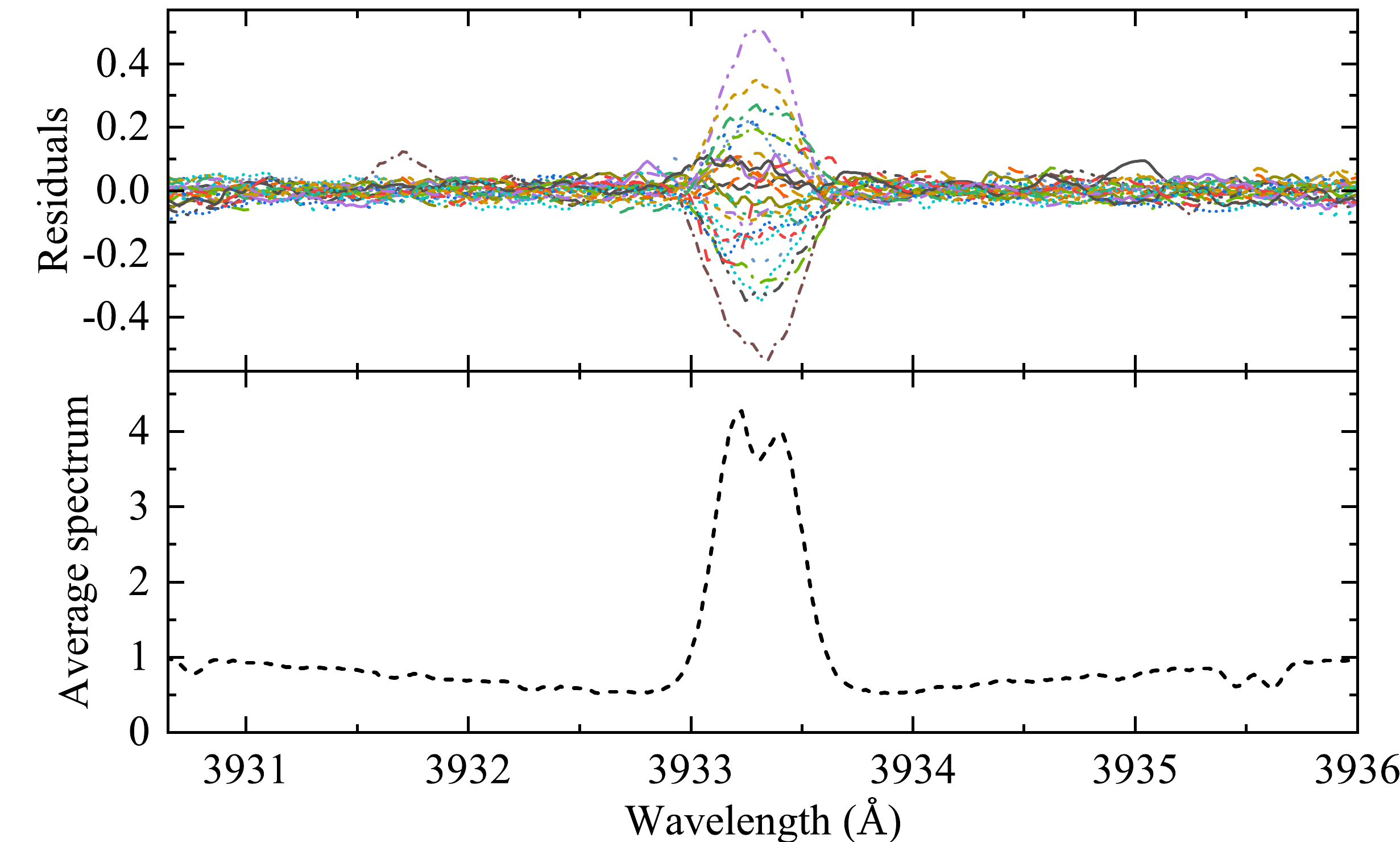}
    \includegraphics[width=\columnwidth]{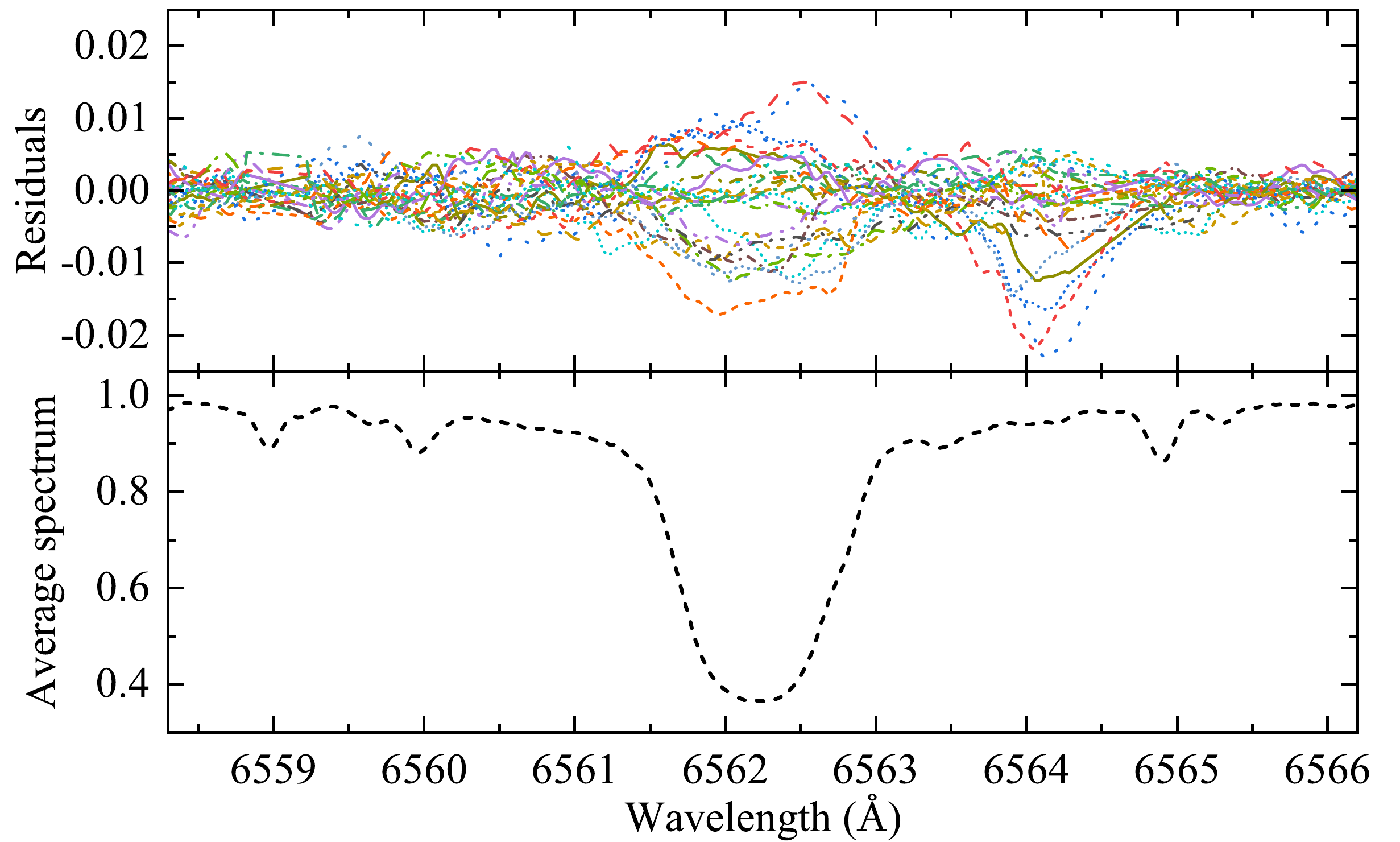}
    \caption{The residual spectrum $f$ (up panel) and the normalized average spectrum $F_M$ (low panal) ($F_\lambda=F_M+f$) for Ca II H (a), K (b) and H$\alpha$ (c).}
    \label{fig:spec_sub_overall}
\end{figure}

Thirdly, each individual spectrum was expanded to two parts: the normalized average spectrum $F_M$ and a residual spectrum $f$,
\begin{equation}
    \label{eq:flux_expand}
    F_\lambda=F_M+f
\end{equation}
Figure \ref{fig:spec_sub_overall} shows the residual spectrum (21-points smoothed) and respective average spectrum of Ca II H \& K and H$\alpha$ observed using HIRES. Note that the fluctuations near 6564 Angstroms are due to the notable water absorption line.

At last EW turned to be
\begin{equation}
    W = W_M+{\Delta}W
    \label{eq:ew_expand}
\end{equation}
Where $W_M$ is the common EW of the normalized average spectrum standing for the absolute activity level of whole dataset, and ${\Delta}W$ is the relative EW of individual spectrum corresponding to the relative variation of chromospheric activity.

For the HIRES spectra, practically $\Delta W$s were calculated by mean of values from integrals over the residual spectra around the centers of the spectral lines (with range of 0.5 Angstroms around Ca II H \& K lines, and range from 6561 to 6563.5 Angstroms for H$\alpha$ line) and from Gaussian function fitting. The measured errors were estimated using the difference between these two methods \citep{Cao2014}.

\begin{figure}
    \includegraphics[width=\columnwidth]{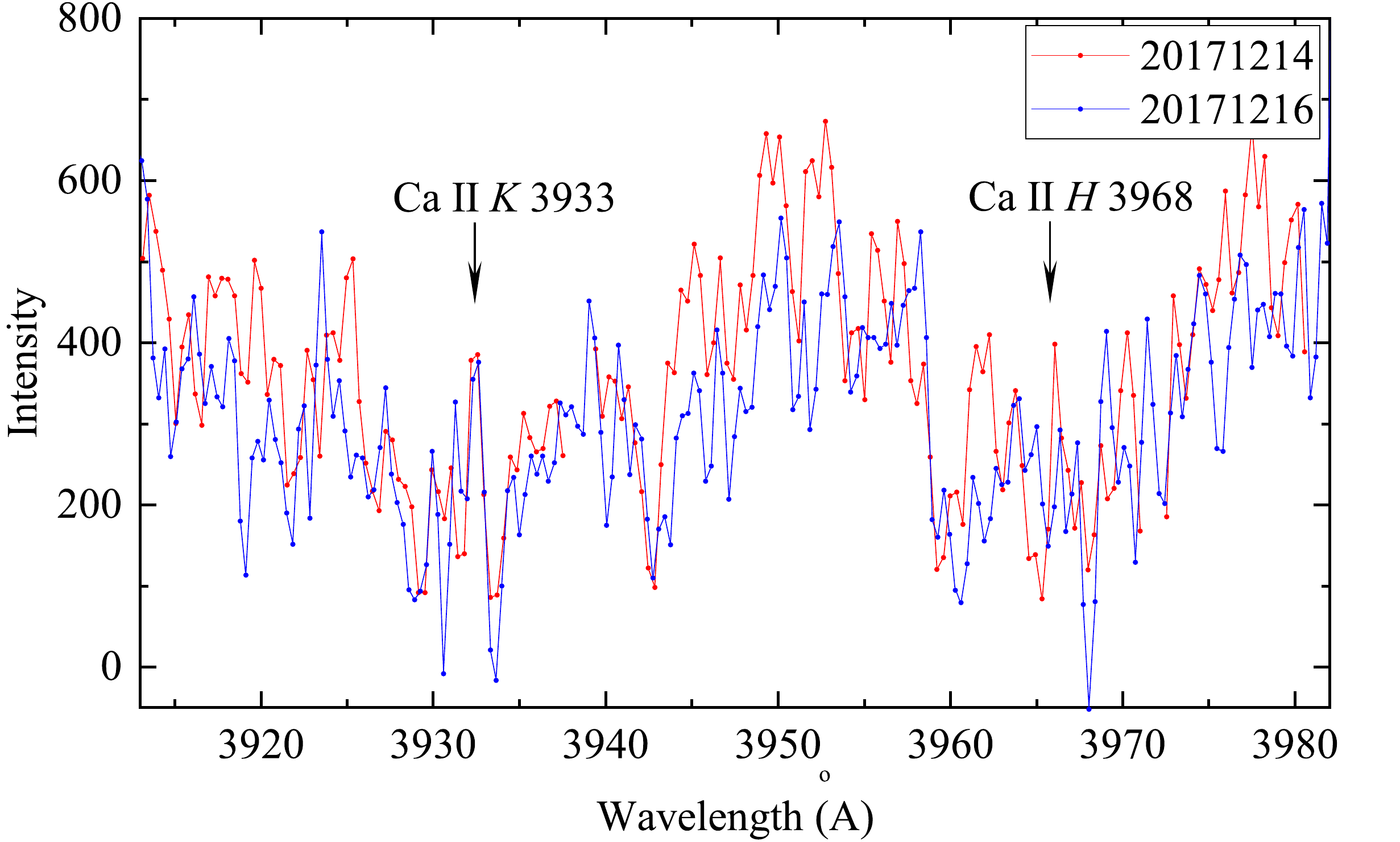}
    \caption{Ca II H \& K lines at Lijiang/YFOSC on Dec 14 and 16, 2017.}
    \label{fig:spec_YFOSC_HK}
\end{figure}

\begin{figure}
    \includegraphics[width=\columnwidth]{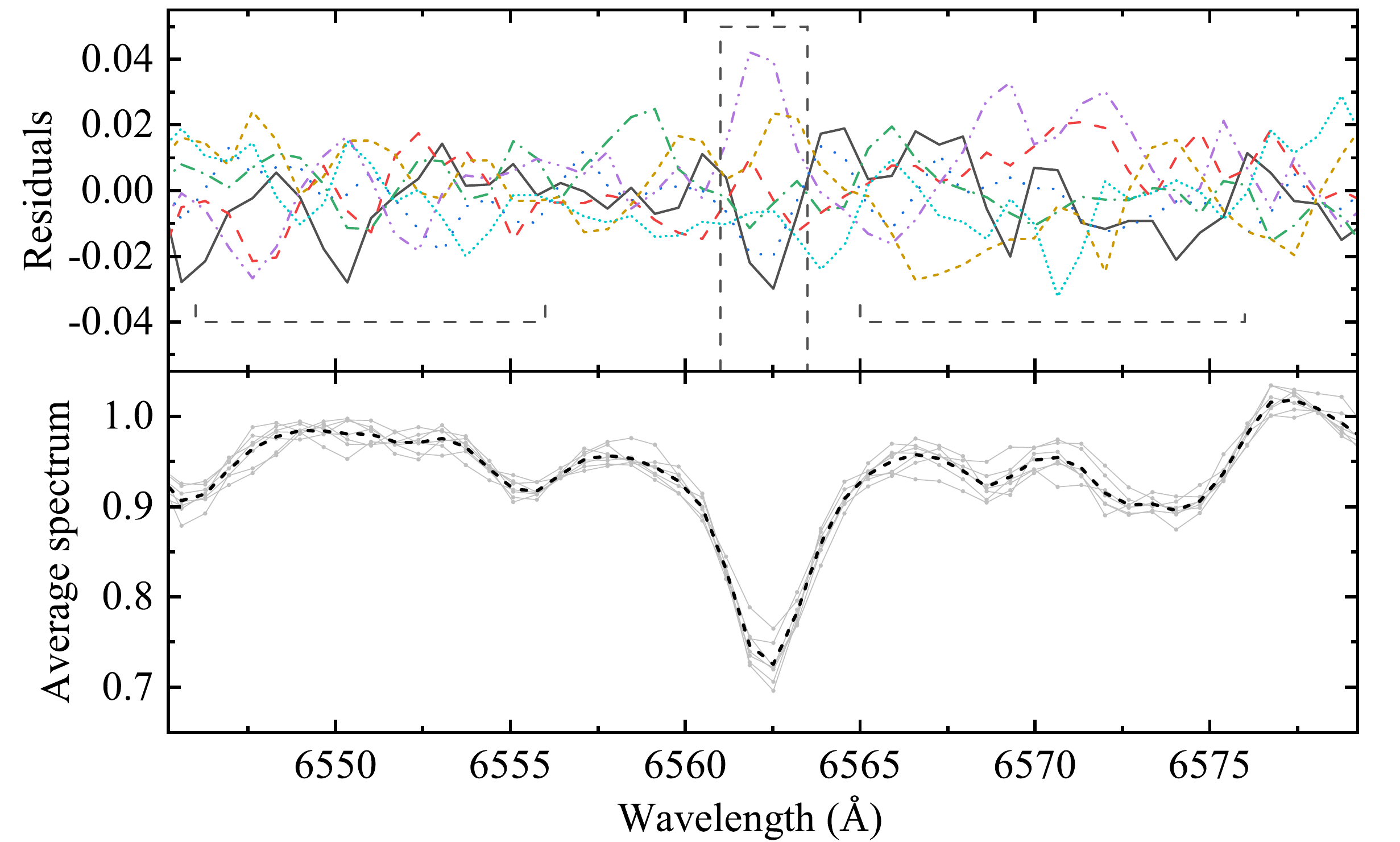}
    \caption{The residuals and average spectrum of H$\alpha$ line at Lijiang/YFOSC. The ranges used to calculate standard error and used to integrate the residual profile are indicated by dash lines.}
    \label{fig:spec_YFOSC_Ha}
\end{figure}

For YFOSC spectra, which are displayed in figures \ref{fig:spec_YFOSC_HK} and \ref{fig:spec_YFOSC_Ha}, from which we can recognize the Ca II H \& K emissions (the other five spectra were ignored due to low SNR) and variation of H$\alpha$ line. The relative EWs of H$\alpha$ line were calculated in the same way as HIRES spectra.

\section{Results and Discussion}\label{sec:rslt}

\subsection{Starspot activity}
\begin{figure}
    \includegraphics[width=\columnwidth]{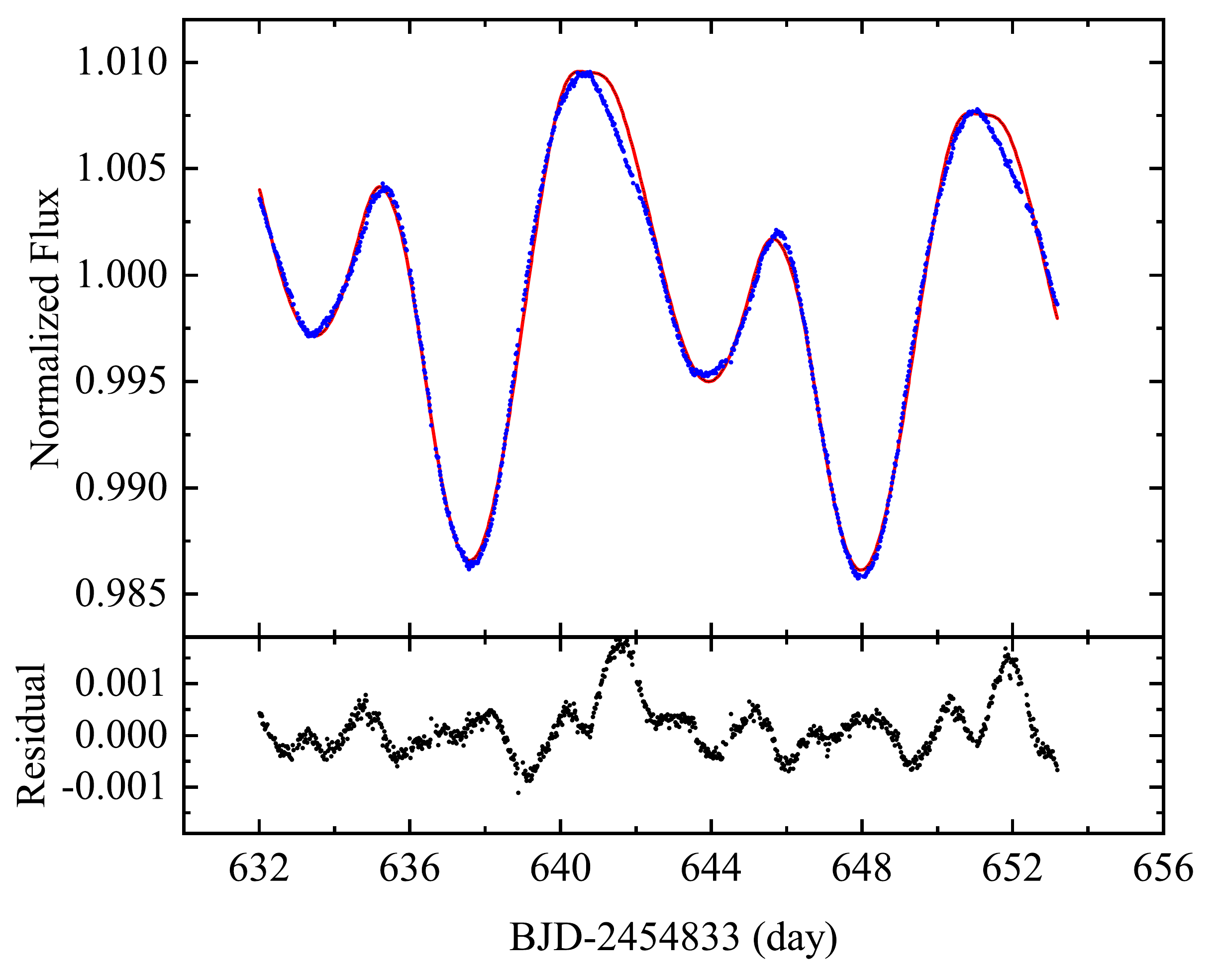}
    \caption{The observed LC (blue dots) and two-spot modeling (red solid line) with fixed latitudes (30$\degr$, 45$\degr$) of the segment at mid-epoch 642.75. The lower panel shows the residuals of fitting. }
    \label{fig:fig_DR_fit_eg}
\end{figure}

The spot configurations at different epochs were inversed under a two-spot LCM with three fixed spot latitude-groups, i.e. (30$\degr$, 45$\degr$), (30$\degr$, 60$\degr$) and (45, 60$\degr$).
As an example, figure \ref{fig:fig_DR_fit_eg} shows fitting of LC segment at mid-epoch 642.75 with latitude-group (30$\degr$, 45$\degr$), from which one can find the signature of SDR: rotation period of 10.382 days for spot at latitude 30$\degr$ (around the primary LC minimum) while 10.456 days for spot at latitude 45$\degr$ (the secondary LC minimum).
Note that occasional distortions, for example, the slight but notable intensity drops around time 642 and 652 in figure \ref{fig:fig_DR_fit_eg}, which can not be explained by such a simple two-spot model, were excluded as a priori before optimization. 
The oscillations in residual with periods of a few days are likely not the effects relating to the stellar rotation, for example the spots, but could be the systematic errors, for example the focus changes due to thermal transients of Kepler telescope \citep[KDCH,][]{KDCH}.

\begin{figure}
    \includegraphics[width=\columnwidth]{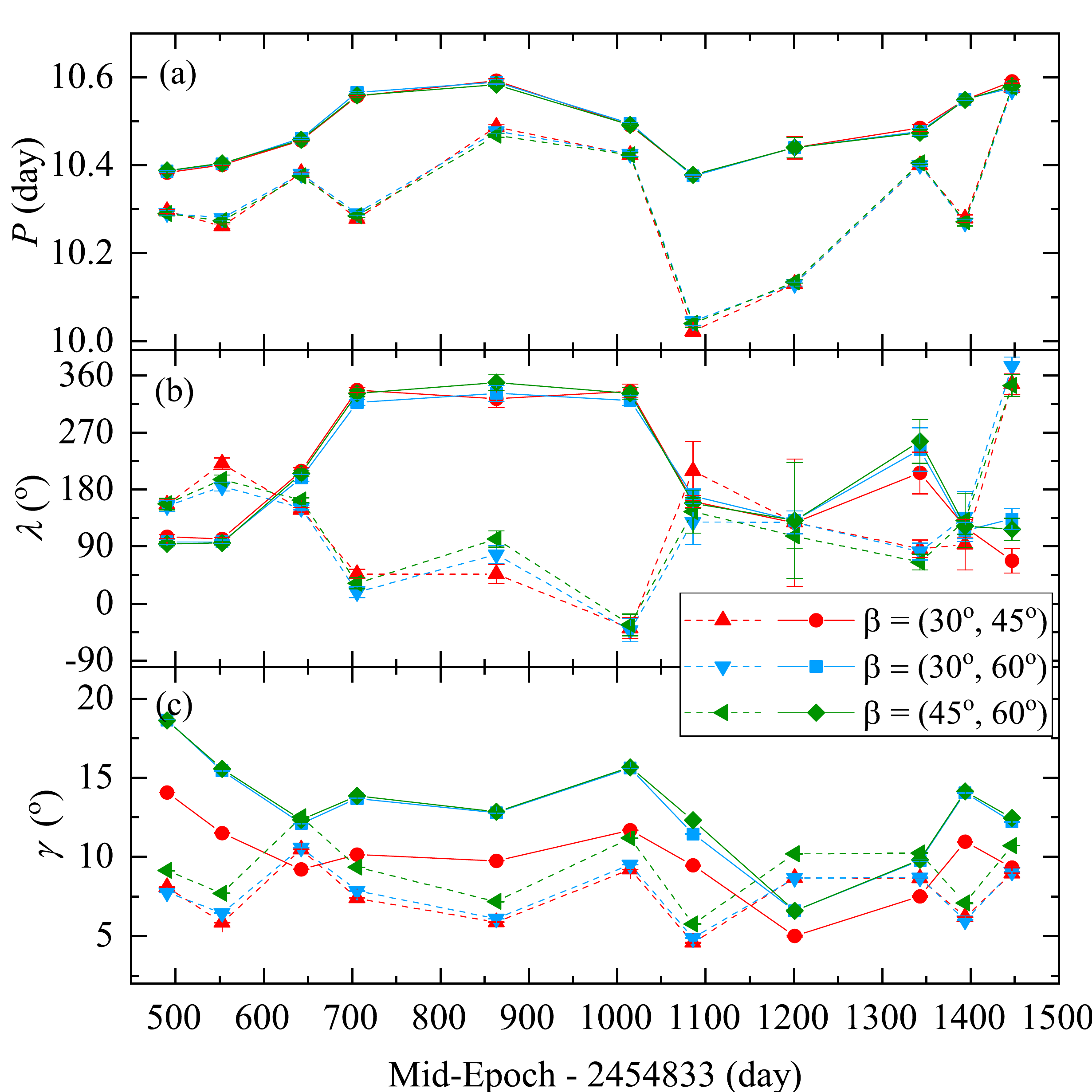}
    \caption{The rotation periods (a), longitudes (b) and radii (c) of two spots from LCM. Values in different epochs are connected by solid and dashed lines for spot at higher and lower latitudes respectively, for convenience. 
The errors (and hereafter) are estimated by error propagation of the observing and spot modeling errors \citep{Press2007}.}
    \label{fig:LCM_combine}
\end{figure}

Figure \ref{fig:LCM_combine} shows the temporal variations of rotation periods, longitudes and radius of two spots in different groups with fixed latitudes. 
It can be seen that rotation periods and longitudes are independent on the input spot-latitudes, except with small deviations which likely come from the different ways of overlaps among individual spots for different choices of spot-latitude groups.

The wide scatter of radius is due to the well-known weak constraints on spot latitude from photometry, i.e. a smaller spot at lower latitude distorts the LC like a larger spot located at higher latitude, especially for a target with extremely high inclination of the rotation axis.
This can be easily found from figure \ref{fig:LCM_combine}(c) by comparing the derived radius of either the spots at higher latitude (hereafter, called "higher" spots, and conversely "lower" spots) between (30$\degr$, 45$\degr$) and (30$\degr$, 60$\degr$) or the "lower" spots between (30$\degr$, 60$\degr$) and (45$\degr$, 60$\degr$).

\subsubsection{SDR}
The existence of multiple rotation periods has been taken as evidence of SDR on stars for a long time. For example, seasonal variations of rotation period along long-term photometry were able to use to detect the lower limit of SDR which depends on how the spot appears closely to the equator and meanwhile highly to the stellar pole within the observing time span \citep{Hall1991, Distefano2016}.

The rotation periods of two spots in figure \ref{fig:LCM_combine}(a) indicate a Solar-like SDR on Kepler-411 because the 
"higher" spots
have relavtively longer rotation period.
Two extreme values can be calculated by means of the periods at mid-epoch 1085.75 and 863.25 respectively: $P_{\text{min}}=10.0361(0.0118)$ days and $P_{\text{max}} = 10.5878(0.0047)$ days, the errors are standard deviation of three input latitude-groups.
Then we can simply estimate the lower limit of the SDR rate by solving Eq.\ref{eq:DR} as
\begin{equation}
    \alpha=(P_{\text{max}}-P_{\text{min}}) / \left(P_{\text{max}}\sin^2(60\degr)-P_{\text{min}}\sin^2(30\degr)\right)
    \label{eq:SDR_rate}
\end{equation}
Where the lower- and higher- limit of spot latitudes come from "thin-flux-tube-evolution model" simulation \citep{Granzer2002} as described in section \ref{sec:model}. It gives $\alpha = 0.1016(0.0023)$ and corresponding equatorial rotation period is $P_{\text{eq}} = 9.7810(0.0169)$ days.

Many great efforts have been devoted to investigate the relation between shear and the stellar parameters for a long time, and it was found that SDR is a function both of effective temperature and rotation rate of the stars \citep[etc., and reference therein]{Hall1991, Henry1995, Barnes2005, Reinhold2015}.
Recently \citet{Balona2016} calculated SDR of 2562 Kepler single stars with spectral type F, G, K and A to measure the frequency spread around the rotation period which provides a lower limit of the rotational shear, and concluded that the SDR rate depends strongly on effective temperature, which is compatible with theoretical models, while its dependence on the rotation rate is a little more complicate, i.e. is weak in K and G stars, increases rapidly in F stars and is the strongest in A stars.
Besides \citet{Reinhold2015} analysed 24124 Kepler stars based on the Lomb-Scargle periodogram, and concluded differently that the SDR rate increases rapidly with increasing of effective temperature between 6200 and 6800 K and otherwise increases slowly.

Despite the relative wide spread of results, it can be assumed from above studies that the expected strength of the shear of SDR for Kepler-411 is about $0.045$ \citep[figure 12]{Reinhold2015} and $0.075 - 0.1$ \citep[figure 4]{Balona2016}, with a effective temperature of $4900 - 5000$ K and a rotation period of about 10 days. 
Note that a value of $\alpha = 0.0521$ can be obtained if we alternatively use a general estimation of $\alpha = (P_{\text{max}} - P_{\text{min}}) / P_{\text{max}}$, i.e. without the constraint from model simulation.
It is clear that both values are familiar with previous studies.

\subsubsection{SDR or spot latitude evolution}

Figure \ref{fig:LCM_combine}(a) displays a quasi-periodic evolution of the rotation period.
From a simple mutation of Eq.\ref{eq:DR} like $\alpha \text{sin}^2(\beta) = 1 - P_{\text{eq}}/P$ one can infer mathematically that this kind of variation can be contributed by the evolution of either spot latitude $\beta$ or SDR rate $\alpha$, or both.

Long-term study on the rapidly rotating young K dwarf AB Dor in terms of the spots \citep{Jeffers2007} revealed that not only the spot configuration, but also DR itself can undergo a temporal cyclical variation, which can be explained as the periodic exchange between kinetic and magnetic energy of star \citep{Petit2002}. This leads to the cyclical variation of the stellar rotation. However there is currently no additional investigation found on this topics.
Due to the lack of constraint on spot latitude, here we do not try to discuss this topic but only note as follows: 1. we can not confirm or deny the existence of variable SDR from the result without additional constraint on spot latitude; 2. there exists latitudinal migration otherwise the evolution of the two spots should be similar to each other, which is practically not the case from result.

\begin{figure}
    \includegraphics[width=\columnwidth]{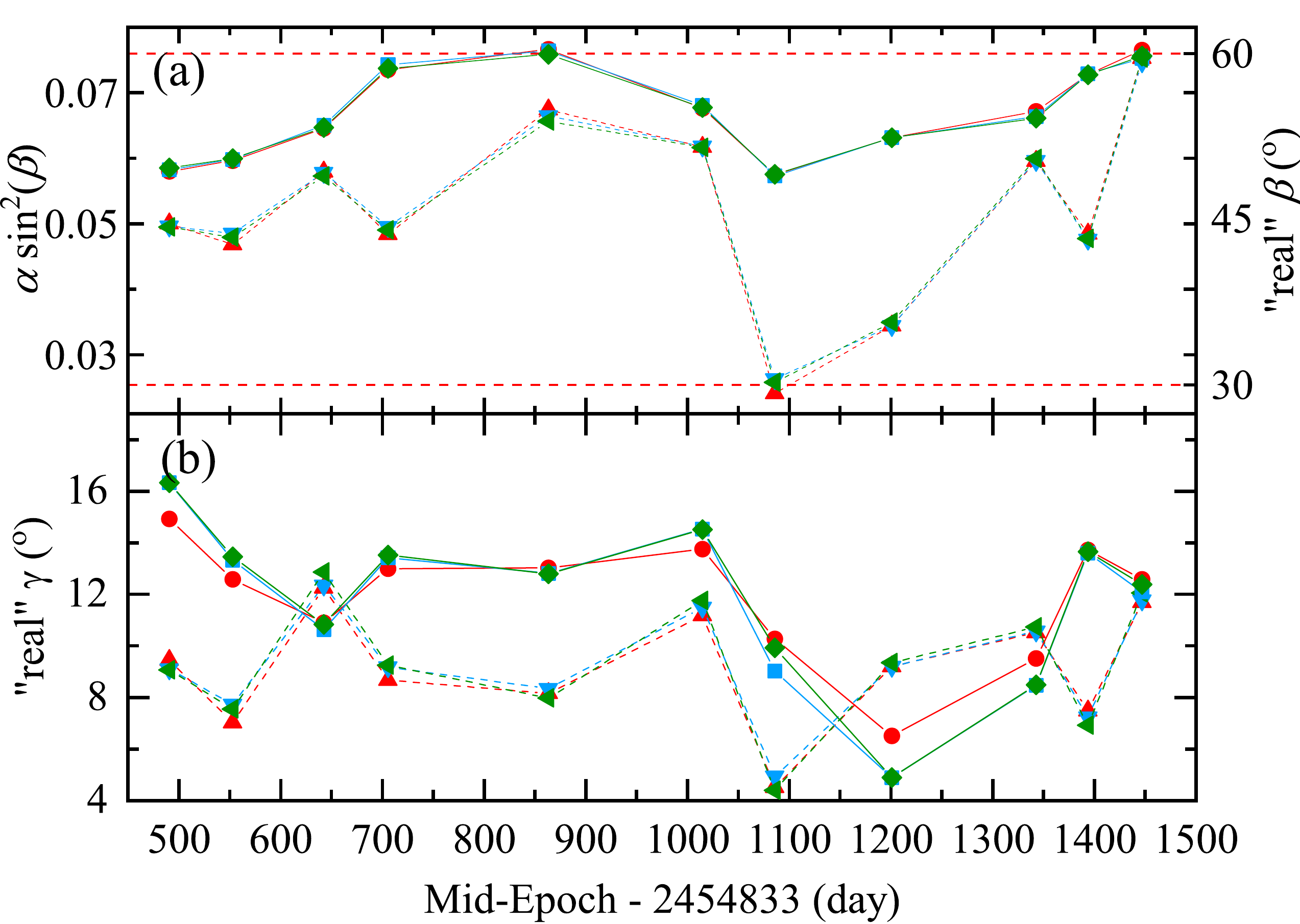}
    \caption{(a) $\alpha \text{sin}^2(\beta)=1-P_{\text{eq}}/P$, the equatorial rotation period $P_{\text{eq}}=9.7810$ days, the right vertical axis gives the "real" latitude $\beta$ calculated with $\alpha=0.1016$.
	(b) The "real" radii of spots.
	The markers are same as figure \ref{fig:LCM_combine}}.
    \label{fig:DR_asinb2_b_rad_cor}
\end{figure}

When spot migrates to or emergences at higher latitude with Solar-like SDR, it plays as a tracer of the rotation of local area and will have a longer period. Meanwhile we can inverse the migration of spot latitude by the variation of measured rotation period when the SDR is known.
Figure \ref{fig:DR_asinb2_b_rad_cor}(a) gives the variation of $\alpha \text{sin}^2(\beta)$, from which we can find the cyclical variation of the spot rotation period. Under the assumption of fixed SDR, we can calculate the latitudinal migration of spots (let's just call it "real" latitude), as labelled at the right vertical axis, calculated by using the value derived in Eq. \ref{eq:SDR_rate}.
The "higher" spot has a good periodic migration of latitude from 45$\degr$ to 60$\degr$, while the "lower" spot migrates with a much larger latitude variation from 30$\degr$ to 60$\degr$.

Then we can further calculate the spots' "real" radii. Considering the fact that spot latitude is usually under a weak constraint in photometry and that a larger spot at higher latitude generates quite similar distortion as a smaller spot at lower latitude, the inversed radius of spot is highly relative to the input latitude.
For simplicity we assume a linear dependence for the increment of the inversed radius to increasing input latitude, i.e. $\Delta\gamma = k\Delta\beta$. The ratio $k$ is decided by fitting $\Delta\gamma-\Delta\beta$ from figure \ref{fig:LCM_combine}(c), where $\Delta\beta = 15\degr$, which gives $k=1.396$ for the "lower" spot and $k=3.21$ for the "higher" spot.
Then the "real" radius can be obtained by $\gamma + k(\beta' - \beta)$, where $\gamma$ means radius corresponding to input latitude $\beta$ as figure \ref{fig:LCM_combine}(c), and $\beta'$ is the "real" latitude as figure \ref{fig:DR_asinb2_b_rad_cor}(a).

The possible diffusion of decaying spots could give rise to confusion in measuring SDR. However, \citet{Giles2017} investigated M- to F-type stars using Kepler data and found that the lifetime of starspots varies with spot sizes, stellar rotation and spectral type, despite the scatter of the results, one could expect a spot lifetime with the value of tens 
of days for the target like ours. 
Thus it is reasonable for us to assume long-lived spots with slightly and linearly evolving radii for each LC segment during the calculation.

\subsubsection{Spot variation}

From figures \ref{fig:LCM_combine} and \ref{fig:DR_asinb2_b_rad_cor} we can see a quasi-periodic evolution of the spot active regions, which can be roughly divided to four durations:
Firstly (from the onset to mid-epoch 650), the "higher" spots has a slightly smaller longitude than "lower" spot, and its radius decreases rapidly, which may imply a relative rapid evolution.
After that the "higher" and "lower" spots migrate longitudinally along different directions, quickly to another semisphere, around zero-longitude, then remain there until mid-epoch 1050. Meanwhile their radii also remain large and vary not much. From that we may infer that the active regions are relatively large and stable within this duration.
From mid-epoch 1050 to 1350, we can see that the spots appear at $\lambda \sim 135\degr$ which is close to the position in the first duration, and remain a small scale of size, and say that the spot activity falls down to low level. 
Lastly (from mid-epoch 1350 to the last LC segment) the spots alternately migrate to longitudes near zero, i.e. another side, and resume to larger size.

Another variable, $\alpha\ \text{sin}^2(\beta)$, also undergoes a highly identical fluctuation as the longitude and radius. This indicates a periodic migration of the spot latitude under assumption of the fixed SDR, or a periodic fluctuation of the SDR rate under assumption of fixed latitude.

From these phenomena we might conclude that the activity on Kepler-411 undergoes a cyclical evolution in terms of spot distribution and size with a time scale of about $660$ days.

\subsection{Chromospheric activity}

The Ca II H \& K core emissions are used to be indicators of magnetic activity. Meanwhile the H$\alpha$ line shows emission above the continuum in very active stars, while appears as a filled-in absorption profile in less active ones \citep{Montes1997}.

Figure \ref{fig:spec_sub_overall} shows that the Ca II H \& K lines have apparent core emissions while H$\alpha$ line appears as a filled-in profile, indicating Kepler-411 as a chromospheric but not very active star.
Observations at YFOSC, see figure \ref{fig:spec_YFOSC_HK}, also show recognizable Ca II H \& K emissions, although with low SNR.
The evolutionary emissions are notable and the relative EWs can be quantitatively measured by using the method described in section \ref{sec:model}.

\begin{figure}
    \includegraphics[width=\columnwidth]{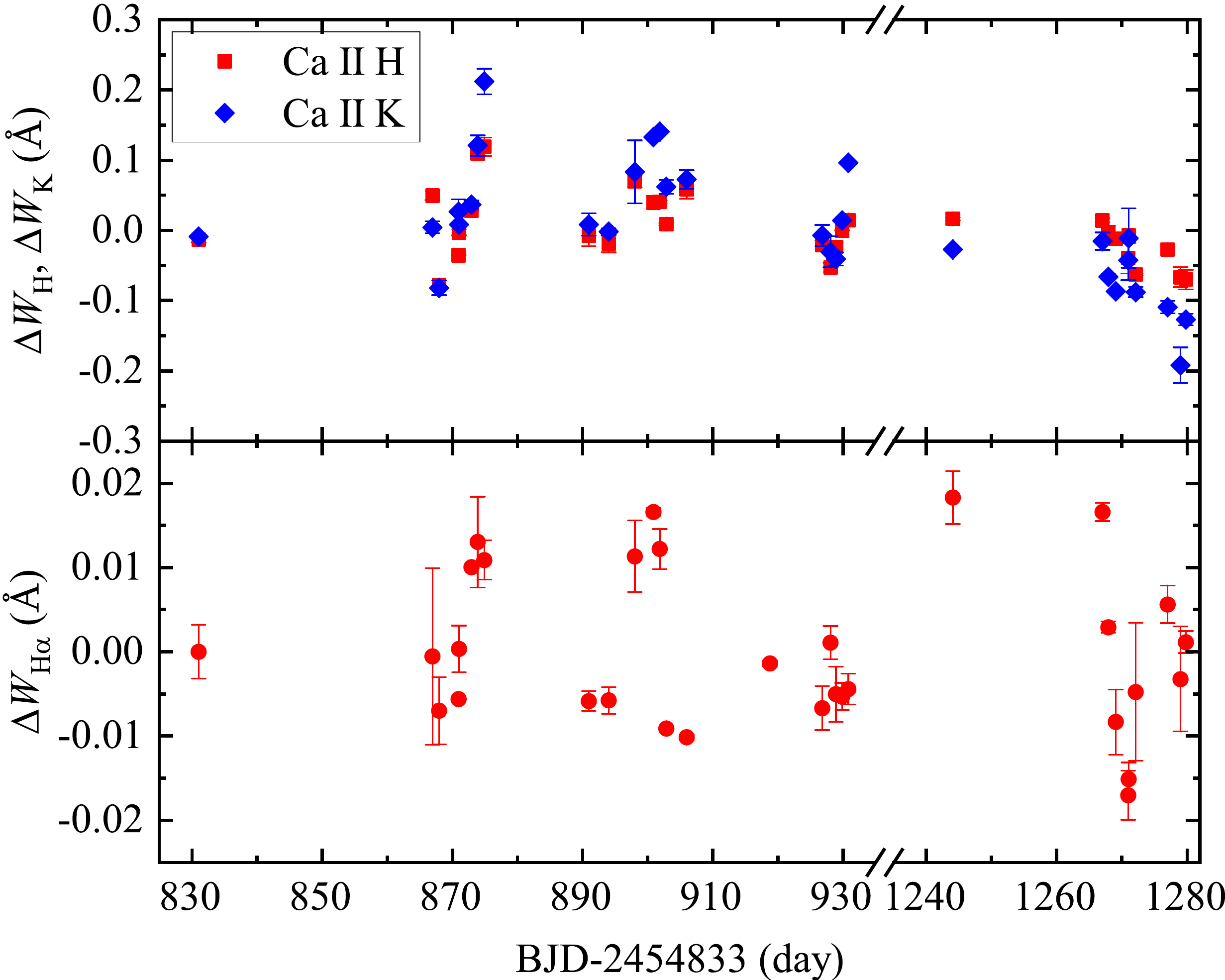}
    \caption{Relative EW versus time for Ca II H \& K lines (upper panel) and H$\alpha$ line (lower panel) of HIRES spectra.}
    \label{fig:EW_HIRES}
\end{figure}

For HIRES spectra, the relative EWs are listed in table \ref{tab:EW_result} and plotted in figure \ref{fig:EW_HIRES}.
We can see that Ca II H \& K lines have similar long-term variations to each other, so does H$\alpha$ line variation but with about 10 times stronger.
\begin{table*}
    \caption{The EWs relative to the background spectra $F_M$, which were taken as the average of all observed spectra.
	The time is BJD at mid-epoch of the exposure.}
    \label{tab:EW_result}
    \begin{tabular}{rrrrrl}
        \hline
		Date     &  Mid epoch  &  $\Delta W_\text{K}$  &   $\Delta W_\text{H}$  & $\Delta W_{\text{H}\alpha}$ & Instrument \\
		yyyymmdd & BJD-2454833 &   Angstrom          &       Angstrom       &    Angstrom           &             \\
		\hline
		20110412 &   831.0313  & -0.0091$\pm$0.0014  &  -0.0129$\pm$0.0041  &  -0.0000$\pm$0.0032   &  Keck/HIRES   \\
		20110518 &   866.9252  &  0.0043$\pm$0.0084  &   0.0494$\pm$0.0066  &  -0.0005$\pm$0.0105   &  Keck/HIRES   \\
		20110519 &   867.9548  & -0.0819$\pm$0.0105  &  -0.0772$\pm$0.0057  &  -0.0070$\pm$0.0040   &  Keck/HIRES   \\
		20110522 &   870.9383  &  0.0265$\pm$0.0180  &  -0.0352$\pm$0.0009  &  -0.0056$\pm$0.0000   &  Keck/HIRES   \\
		20110522 &   871.0228  &  0.0082$\pm$0.0035  &  -0.0027$\pm$0.0046  &   0.0003$\pm$0.0027   &  Keck/HIRES   \\
		20110524 &   872.9298  &  0.0367$\pm$0.0058  &   0.0282$\pm$0.0008  &   0.0100$\pm$0.0002   &  Keck/HIRES   \\
		20110525 &   873.9056  &  0.1208$\pm$0.0143  &   0.1091$\pm$0.0048  &   0.0130$\pm$0.0054   &  Keck/HIRES   \\
		20110526 &   874.9073  &  0.2121$\pm$0.0183  &   0.1192$\pm$0.0129  &   0.0109$\pm$0.0023   &  Keck/HIRES   \\
		20110611 &   890.9330  &  0.0084$\pm$0.0158  &  -0.0079$\pm$0.0143  &  -0.0058$\pm$0.0012   &  Keck/HIRES   \\
		20110614 &   894.0393  & -0.0021$\pm$0.0048  &  -0.0189$\pm$0.0125  &  -0.0058$\pm$0.0016   &  Keck/HIRES   \\
		20110618 &   898.0297  &  0.0833$\pm$0.0449  &   0.0696$\pm$0.0079  &   0.0113$\pm$0.0043   &  Keck/HIRES   \\
		20110621 &   900.9030  &  0.1329$\pm$0.0012  &   0.0397$\pm$0.0093  &   0.0166$\pm$0.0005   &  Keck/HIRES   \\
		20110622 &   901.8542  &  0.1404$\pm$0.0030  &   0.0407$\pm$0.0020  &   0.0122$\pm$0.0024   &  Keck/HIRES   \\
		20110623 &   902.8281  &  0.0621$\pm$0.0100  &   0.0088$\pm$0.0019  &  -0.0091$\pm$0.0001   &  Keck/HIRES   \\
		20110626 &   905.9932  &  0.0724$\pm$0.0130  &   0.0585$\pm$0.0137  &  -0.0101$\pm$0.0004   &  Keck/HIRES   \\
		20110709 &   918.7891  &  ...                &   ...                &  -0.0014$\pm$0.0004   &  Keck/HIRES   \\
		20110717 &   926.7929  & -0.0073$\pm$0.0151  &  -0.0207$\pm$0.0061  &  -0.0067$\pm$0.0026   &  Keck/HIRES   \\
		20110718 &   928.0932  & -0.0304$\pm$0.0222  &  -0.0525$\pm$0.0066  &   0.0011$\pm$0.0020   &  Keck/HIRES   \\
		20110719 &   928.8968  & -0.0406$\pm$0.0094  &  -0.0235$\pm$0.0070  &  -0.0050$\pm$0.0033   &  Keck/HIRES   \\
		20110720 &   929.8823  &  0.0140$\pm$0.0014  &  -0.0003$\pm$0.0001  &  -0.0053$\pm$0.0016   &  Keck/HIRES   \\
		20110721 &   930.7977  &  0.0960$\pm$0.0023  &   0.0145$\pm$0.0050  &  -0.0044$\pm$0.0018   &  Keck/HIRES   \\
		20120529 &  1244.0810  & -0.0272$\pm$0.0012  &   0.0166$\pm$0.0022  &   0.0183$\pm$0.0032   &  Keck/HIRES   \\
		20120621 &  1267.0274  & -0.0154$\pm$0.0125  &   0.0144$\pm$0.0028  &   0.0166$\pm$0.0011   &  Keck/HIRES   \\
		20120622 &  1267.9237  & -0.0659$\pm$0.0037  &  -0.0043$\pm$0.0109  &   0.0029$\pm$0.0007   &  Keck/HIRES   \\
		20120623 &  1269.1029  & -0.0867$\pm$0.0009  &  -0.0116$\pm$0.0019  &  -0.0083$\pm$0.0039   &  Keck/HIRES   \\
		20120625 &  1271.0143  & -0.0423$\pm$0.0287  &  -0.0397$\pm$0.0217  &  -0.0170$\pm$0.0029   &  Keck/HIRES   \\
		20120625 &  1271.0321  & -0.0110$\pm$0.0422  &  -0.0072$\pm$0.0051  &  -0.0151$\pm$0.0020   &  Keck/HIRES   \\
		20120626 &  1272.1107  & -0.0877$\pm$0.0076  &  -0.0628$\pm$0.0013  &  -0.0048$\pm$0.0082   &  Keck/HIRES   \\
		20120701 &  1277.0206  & -0.1094$\pm$0.0088  &  -0.0275$\pm$0.0051  &   0.0056$\pm$0.0022   &  Keck/HIRES   \\
		20120703 &  1278.9877  & -0.1919$\pm$0.0256  &  -0.0669$\pm$0.0142  &  -0.0032$\pm$0.0062   &  Keck/HIRES   \\
		20120704 &  1279.8399  & -0.1269$\pm$0.0082  &  -0.0699$\pm$0.0139  &   0.0011$\pm$0.0013   &  Keck/HIRES   \\
		20171214 &  3269.0232  &  ...                &  ...                 &   0.0363$\pm$0.0123   &  Ljiang/YFOSC   \\
		20171215 &  3270.0303  &  ...                &  ...                 &   0.0008$\pm$0.0121   &  Ljiang/YFOSC   \\
		20171216 &  3270.9987  &  ...                &  ...                 &   0.0294$\pm$0.0079   &  Ljiang/YFOSC   \\
		20171217 &  3272.0029  &  ...                &  ...                 &   0.0100$\pm$0.0083   &  Ljiang/YFOSC   \\
		20171223 &  3278.0090  &  ...                &  ...                 &  -0.0641$\pm$0.0160   &  Ljiang/YFOSC   \\
		20171228 &  3283.0023  &  ...                &  ...                 &  -0.0296$\pm$0.0142   &  Ljiang/YFOSC   \\
		20171229 &  3284.0024  &  ...                &  ...                 &   0.0172$\pm$0.0109   &  Ljiang/YFOSC   \\
        \hline
    \end{tabular}
\end{table*}
\begin{figure}
    \includegraphics[width=\columnwidth]{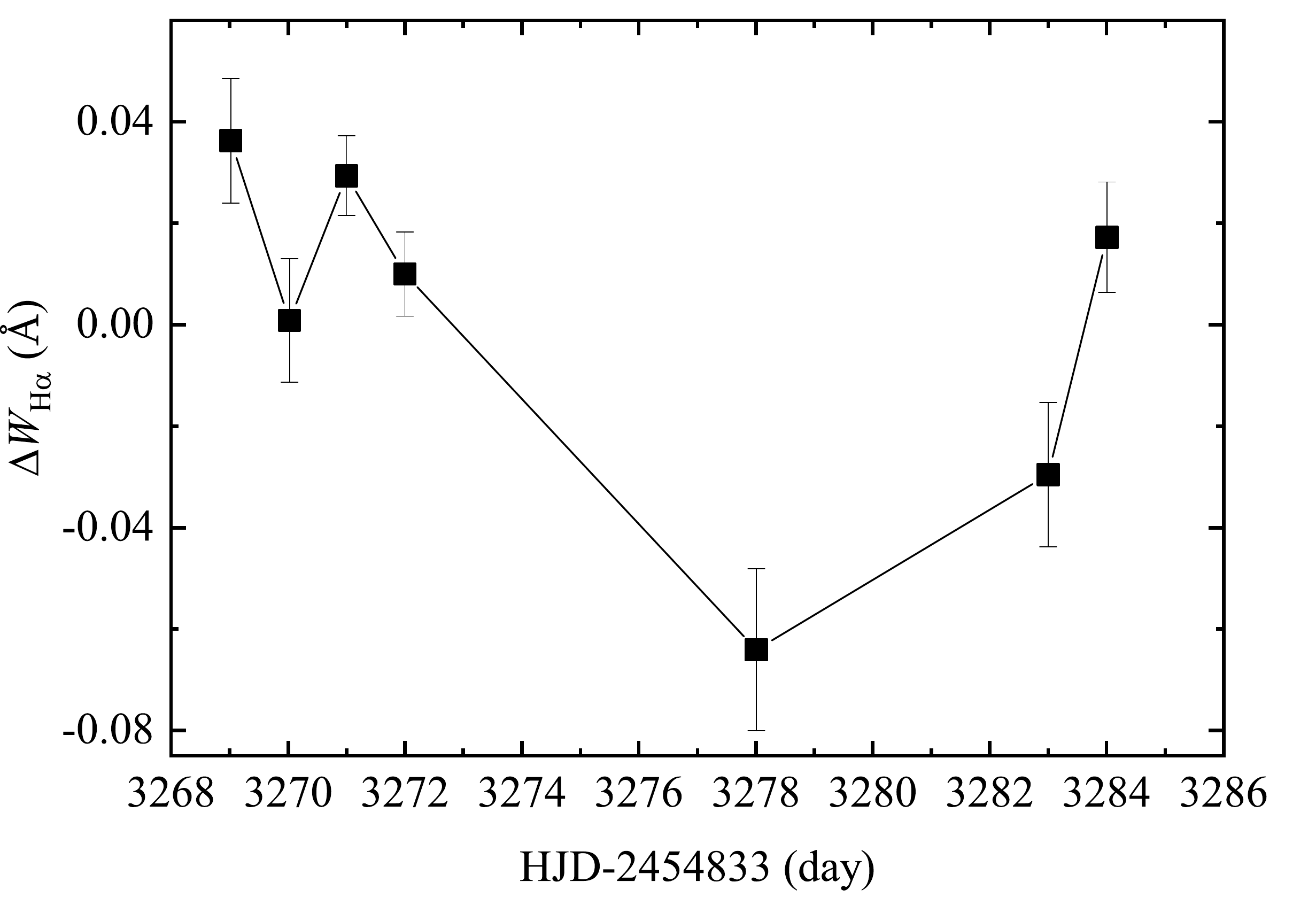}
    \caption{Relative EW versus time for H$\alpha$ line of YFOSC spectra.}
    \label{fig:EW_YFOSC}
\end{figure}

The observations at YFOSC span cover over one rotation period, the H$\alpha$ line shows much larger variation than the observations in 2011 and 2012, see table \ref{tab:EW_result} and figure \ref{fig:EW_YFOSC}.

\subsubsection{Correlation between indicators}

It was found that the Ca II H and K excess emissions are strongly correlated with each other 
and practically they are usually measured together to estimate the stellar activity, for example, the wide-used S-index \citep{Vaughan1978}.
On the other hand, correlations of H$\alpha$ with other diagnostics are more complicated, because with increasing heating rate, H$\alpha$ first becomes a deeper absorption line and then goes into emission. This nonlinear behavior has been predicted by models and proven by observations \citep{Linsky2017}.

\begin{figure}
    \includegraphics[width=\columnwidth]{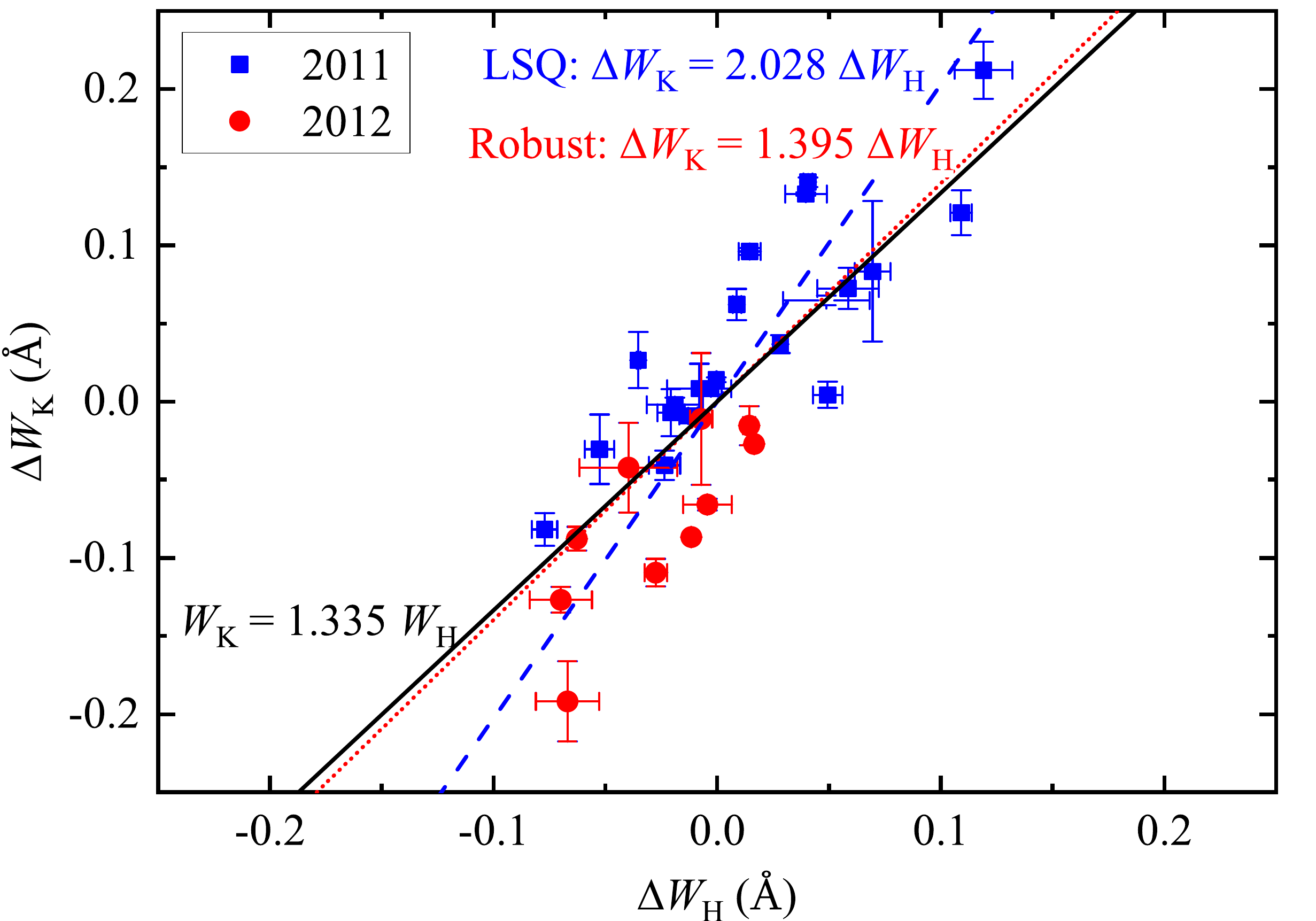}
    \caption{Relative EW of Ca II K line versus Ca II H line, the ratio of $W_\text{K}/W_\text{H}$ and linear fits using the least-square-quadratic (LSQ) and robust regressing methods are over-plotted. }
    \label{fig:EW_relation_K_H}
\end{figure}

Figure \ref{fig:EW_relation_K_H} shows the correlation between relative EWs of Ca II K and H core emissions. We can see that the relative increase of Ca II K line is strongly correlated with Ca II H line variation.
Note that the ratio between them is not $1$.
A linear fit to the ratio between relative EWs of them gives $2.028$ using least-square-quadratic (LSQ) and $1.395$ using robust regressing method. The later is very close to the ratio between the EWs of the Ca II H and K lines, $W_\text{K}/W_\text{H} = 1.335$, which are obtained by directly integrating on a 1.1 Angstroms width centered on the emission cores of corresponding average spectra as shown in figure \ref{fig:spec_sub_overall}, $W_\text{H} = -1.611 A$ and $W_\text{K} = -2.151 A$ respectively.
Furthermore, we can also see a notable difference of such relation between 2011 and 2012, although with wide scatter, i.e. the ratio between increases of Ca II K and H emissions is relatively larger in 2012 than 2011.

\begin{figure}
    \includegraphics[width=\columnwidth]{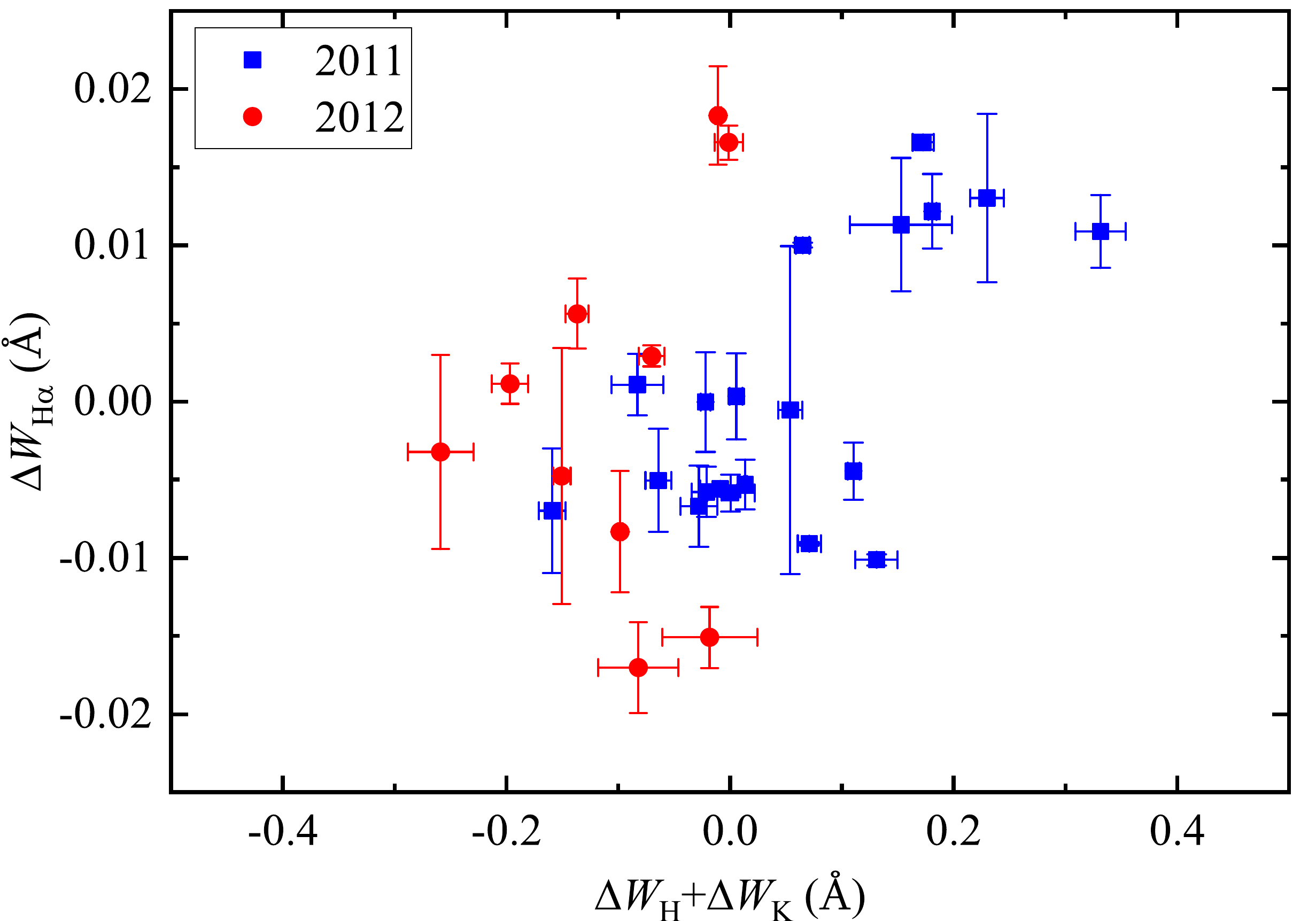}
    \caption{Relative EW of H$\alpha$ line versus Ca II H+K lines. The markers are used same as figure \ref{fig:EW_relation_K_H}.}
    \label{fig:EW_relation_Ha_HK}
\end{figure}

A positive correlation between H$\alpha$ and Ca II emissions can also be roughly recognized from figure \ref{fig:EW_relation_Ha_HK}, although with wide scatter.
We can find again the difference between 2011 and 2012. In 2011 Ca II H \& K lines show wide range of activity levels with both weaker and stronger emissions than average spectrum and H$\alpha$ line's relative EWs appear a positive correlation with Ca II H \& K lines. While in 2012, when Ca II H \& K lines show stronger emissions in all spectra, H$\alpha$ line obeys no longer the positive correlation as before, and there are even a few points have apparently the weakest emissions.

\subsubsection{Connection between photospheric and chromospheric activities}
A correlation between photometric variation and chromospheric emission would indicate a co-located connection of the magnetic field spreading over photosphere and chromosphere.
However both positive and negative examples were found in previous studies.
For example, a large dark sunspot in the solar photosphere is almost always spatially associated with a bright plage while small plages often appear without cospatial sunspots \citep{Ash2020}. 
\citet{Vida2015} found a correlation between H$\alpha$ line and LC under rotational modulation of FK Comae. 
\citet{Flores-Soriano2017} observed several active regions on LQ Hydrae but only one of them shows significant changes in its chromospheric H$\alpha$ emission. 
\citet{Morris2018} found both examples with and without correlation between S-index and starspots, 
and \citet{Kriskovics2019} found no clear correlation between the position of the strongest chromospheric activity and the biggest spot coverage on V1358 Orinis. 
\citet{Ash2020} further found opposite way of H$\alpha$ emission w.r.t. flux and concluded starspot-dominated and plage-driven magnetic variabilities on K and M dwarfs.

\begin{figure}
    \includegraphics[width=\columnwidth]{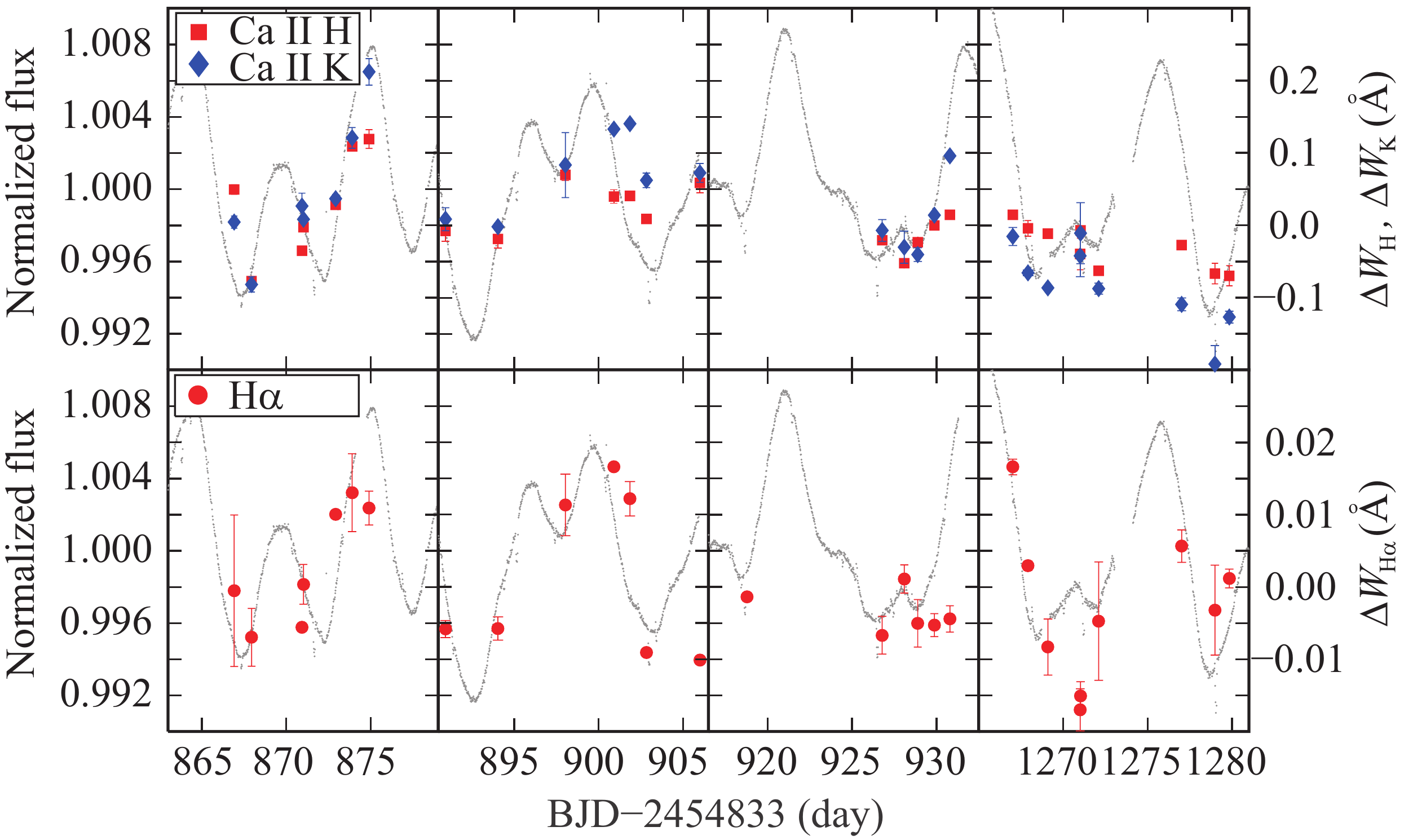}
    \caption{Rotational modulations of relative EW of the Ca II H \& K lines (upper panels) and H$\alpha$ line (lower panels) at HIRES compared with photometric normalized flux from Kepler during 4 segments of 16 days.
        The LCs are plotted with gray dots, the relative EWs are over-plotted with error bars: the red squares and blue diamonds in the upper panels represent the relative EWs of Ca II H and K lines, respectively, while the red solid circles in the lower panels represent the relative EWs of H$\alpha$.
    }
    \label{fig:EW_LC}
\end{figure}

From figure \ref{fig:EW_HIRES} one can apparently find the variabilities of the spectral lines at different epochs, which vary a little large in short time-scale of several days and even much notable than the long-time variations.
The spectroscopic observations at HIRES were done within 2011 and 2012, which matched to the quarters of Q09, Q10, Q13 and Q14 of Kepler continuous photometry and covered several rotations.
Figure \ref{fig:EW_LC} shows the comparison of the relative EWs of the Ca II H \& K and H$\alpha$ lines with the photometric LC of Kepler within 4 LC segments with time span of 16 days.

\begin{figure}
    \includegraphics[width=\columnwidth]{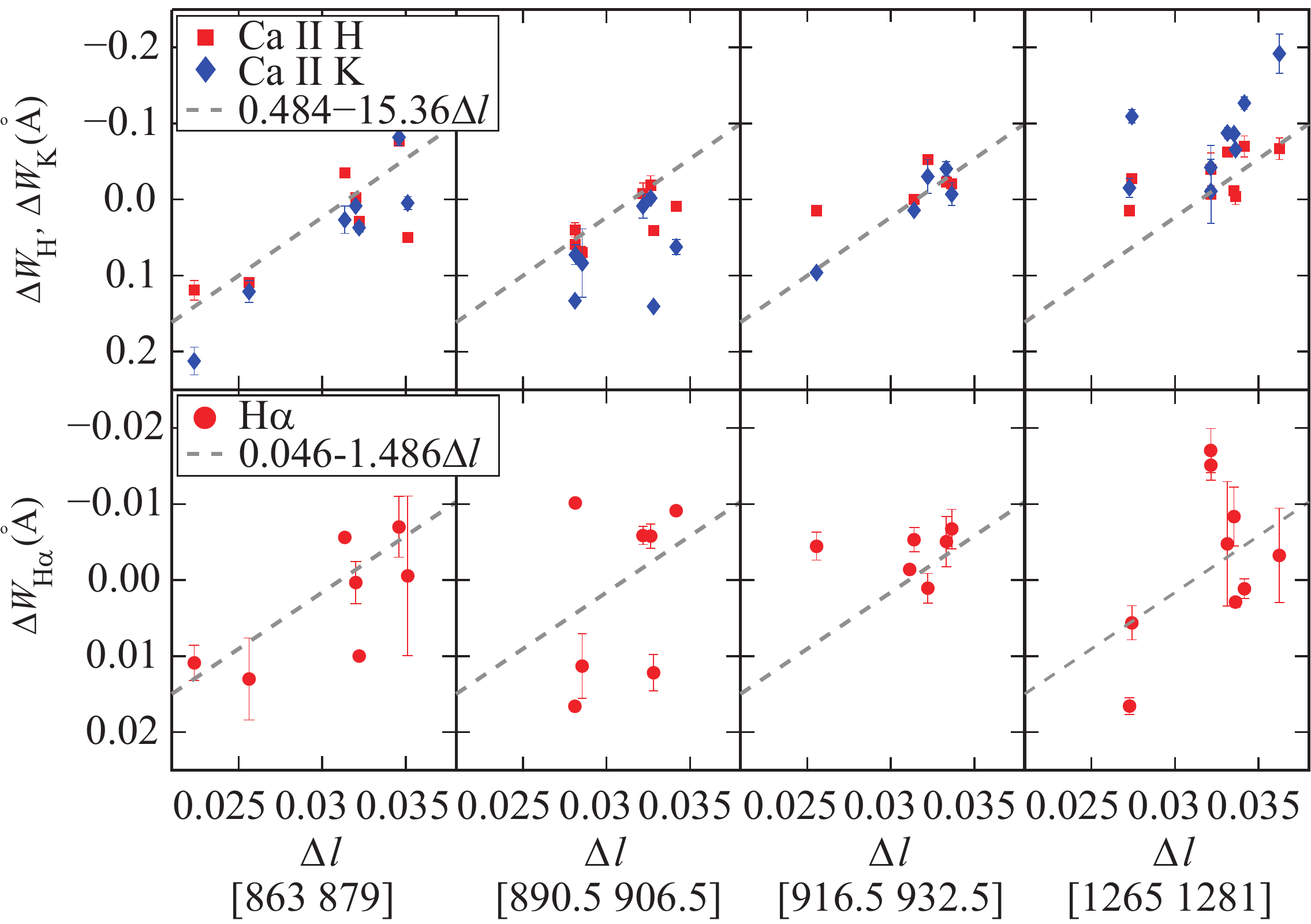}
    \caption{Relative EW versus the dimming of brightness ${\Delta}l$, from measured flux to the maximum flux 1.03, for the Ca II H \& K lines (upper panels) and H$\alpha$ line (lower panels), during 4 segments same to figure \ref{fig:EW_LC}.
		The symbols of $\Delta W$s are same as figure \ref{fig:EW_LC}.
       A robust regression of a linear fit for relative EWs of Ca II H \& K and H$\alpha$ lines, are over-plotted with gray dash lines.}
    \label{fig:EW_dLC}
\end{figure}

To further quantify this correlation, figure \ref{fig:EW_dLC} shows the scatter of the relative EWs versus the flux differences, ${\Delta}l$, between the measured flux and the maximum flux $1.03$.
Besides using measured weights for all Ca II H \& K data points, a robust regression, as mentioned in section \ref{sec:model}, of linear fit yields ${\Delta}W=0.484-15.37{\Delta}l$, while the same relation for H$\alpha$ data points yields ${\Delta}W=0.046-1.49{\Delta}l$, i.e. these two coefficient values are about one-tenth of the case of Ca II H \& K lines, which is reasonable considering that its amplitude of variation is about one-tenth of Ca II H \& K lines, as mentioned above.

It can be evidently seen a tight correlation between the relative EW and variation in LC, i.e., the chromospheric emission approaches to be stronger when the photospheric disk becomes fainter, although the normalization of the two variables is chosen arbitrarily to some extent to scale them together. 
This implies that the periodic dimming of the LC is most likely caused by the appearances of spots 
because the chromospheric emission becomes stronger when the spot appear on its visible hemisphere.
The chromospheric active region is associated with the photospheric starspot region in the spatial structure, and the localized magnetic loop heating the chromosphere is connected to the spot region. 

\section{Conclusions}\label{sec:concl}

By employing the continuous and high-precision photometric data from the space-based Kepler telescope, we have modelled the LC of Kepler-411 at series of epochs with a simplified two-spots model and derived the spot distribution and their temporal evolution. 
Based on the analysis of the high-resolution spectroscopic observations taken at HIRES of Keck telescope during observing runs from 2011 to 2012, and our auxiliary medium-resolution spectroscopic observations at YFOSC of Lijiang 2.4-m telescope, we derived the relative EW variations of the Ca II H \& K and H$\alpha$ lines to analyze its chromospheric activity.
Our main results are summarized as follows:

\begin{enumerate}[1)]
\item
Based on three groups of spots,
the result indicates both a quasi-periodic evolution of spots on timescale of about 660 days and existence of the SDR  with a lower limit of the SDR rate as $\alpha = 0.1016(0.0023)$, and the corresponding equatorial rotation period is $P_{\text{eq}} = 9.7810(0.0169)$ days.

\item
The Ca II H and K excess emissions are strongly correlated with each other, and there also exists a correlation between H$\alpha$ and Ca II H \& K emissions but with large scatter, as pointed out in previous studies.
Furtherly the correlation between Ca II H and K emissions reveals an interesting evolution from year 2011 to 2012.

\item
The chromospheric activity is found to be tightly anti-correlated with photometric variation, i.e. the chromospheric emissions have high coherence with the dimming of the photospheric brightness under the rotational modulation. 
This indicates that the chromospheric active region is associated with the photospheric starspot region, suggesting that the localized magnetic loop heating the chromospheric active region is connected to the spot region.

\end{enumerate}

\section*{Acknowledgments}
We would like to thank the staff of the 2.4-m telescope of Yunnan Observatories for supporting our observations.
Funding for the 2.4 m telescope has been provided by the Chinese Academy of Sciences and the People's Government of Yunnan Province.
This research has made use of the SIMBAD database, operated at CDS, Strasbourg, France, the Exoplanet Orbit Database and the Exoplanet Data Explorer at exoplanets.org.
This research has also made use of the Keck Observatory Archive (KOA), which is operated by the W. M. Keck Observatory and the NASA Exoplanet Science Institute (NExScI), under contract with the National Aeronautics and Space Administration.
All photometric data presented in this paper were obtained from the Mikulsky Archive for Space Telescopes (MAST). STScI is operated by the Association of Universities for Research in Astronomy, Inc., under NASA contract NAS5-26555. 
This paper includes data collected by the Kepler mission. Funding for the Kepler mission is provided by the NASA Science Mission directorate.
This work is supported by National Natural Science Foundation of China through grants Nos. 10373023, 10773027, 11333006 and U1531121.
The joint research project between Yunnan Observatories and Hamburg Observatory is funded by Sino-German Center for Research Promotion (GZ1419).
%
\section*{Data availability statement} 
The data underlying this article are available in the article. 

\bibliographystyle{mnras}
\bibliography{0_title_abs}


\bsp	
\label{lastpage}
\end{document}